\documentclass[]{aastex631}

\usepackage{amsmath}	
\usepackage{amssymb}	
\usepackage[english]{babel}
\usepackage{blindtext}  
\usepackage{booktabs}   
\usepackage{enumitem}   
\usepackage{graphicx}	
\usepackage{longtable}
\usepackage{mhchem}     
\usepackage{multirow}   
\usepackage{pifont}     

\defcitealias{genda2017terrestrial}{G17}
\defcitealias{marchi2018heterogeneous}{M18}
\defcitealias{citron2022large}{C22}

\received{July 25, 2023}
\revised{October 26, 2023}

\submitjournal{PSJ}

\shorttitle{Strength-based simulations of impactor iron distributions}
\shortauthors{Itcovitz et al.}

\graphicspath{{./}{figures/}}

\begin{document}

\title{The distribution of impactor core material during large impacts on Earth-like planets}

\author[0000-0003-2079-8171]{Jonathan P. Itcovitz}
\affiliation{Institute of Astronomy, University of Cambridge, Madingley Road, Cambridge, CB3 0HA, UK}
\email{ji263@cam.ac.uk}

\author{Auriol S.P. Rae}
\affiliation{Department of Earth Sciences, University of Cambridge, Downing Street, Cambridge CB2 3EQ, UK}

\author{Thomas M. Davison}
\affiliation{Impacts and Astromaterials Research Centre, Department of Earth Science and Engineering, Imperial College London, UK}

\author{Gareth S. Collins}
\affiliation{Impacts and Astromaterials Research Centre, Department of Earth Science and Engineering, Imperial College London, UK}

\author{Oliver Shorttle}
\affiliation{Institute of Astronomy, University of Cambridge, Madingley Road, Cambridge, CB3 0HA, UK}
\affiliation{Department of Earth Sciences, University of Cambridge, Downing Street, Cambridge CB2 3EQ, UK}

\begin{abstract}
Large impacts onto young rocky planets may transform their compositions, creating highly reducing conditions at their surfaces and reintroducing highly siderophile metals to their mantles. Key to these processes is the availability of an impactor's chemically reduced core material (metallic iron). It is, therefore, important to constrain how much of an impactor's core remains accessible to a planet's mantle/surface, how much is sequestered to its core, and how much escapes. Here, we present 3D simulations of such impact scenarios using the shock physics code iSALE to determine the fate of impactor iron. iSALE's inclusion of material strength is vital in capturing the behavior of both solid and fluid components of the planet and thus characterizing iron sequestration to the core. We find that the mass fractions of impactor core material that accretes to the planet core ($f_{core}$) or escapes ($f_{esc}$) can be readily parameterized as a function of a modified specific impact energy, with $f_{core} > f_{esc}$ for a wide set of impacts. These results differ from previous works that do not incorporate material strength. Our work shows that large impacts can place substantial reducing impactor core material in the mantles of young rocky planets. Impact-generated reducing atmospheres may thus be common for such worlds. However, through escape and sequestration to a planet's core, large fractions of an impactor's core can be geochemically hidden from a planet's mantle. Consequently, geochemical estimates of late bombardments of planets based on mantle siderophile element abundances may be underestimates.

\end{abstract}


\keywords{Planetary science(1255) --- Impact phenomena(779)}

\section{Introduction} \label{sec:Introduction}
Large impacts can substantially affect the composition and climate of terrestrial planets during their formation and early evolution. Such impacts, potentially also referred to as sub-catastrophic impacts, consist of approximately Ceres- to Pluto-sized projectiles which are sourced from the leftover population of planetesimals following planet formation; for Earth, this means an impactor-to-target mass ration of $\lesssim 0.01$ (e.g., \citealt{bottke2010stochastic, marchi2014widespread}; see Section \ref{sec:param_applicable}). \textit{Large} impact events are significantly less energetic than \textit{giant} impacts (e.g., the Moon-forming impact, \citealt{canup2012forming, cuk2012making, sleep2014terrestrial, lock2018origin}). Most importantly, this means that the target planet's mantle is melted to a lesser extent, and core-mantle re-equilibration is limited. Large impacts can, nonetheless, melt substantial portions of the target mantle \citep{tonks1993magma, pierazzo1997reevaluation, nakajima2021scaling, citron2022large}, and can result in considerable reducing power being accreted to the target. The atmosphere and interior of the target planet thus have the potential to be strongly chemically affected by large impacts, with such scenarios having been suggested as providing globally reducing surface conditions for prebiotic chemistry \citep{benner2020when, zahnle2020creation} and delivering much of Earth's accessible budget of siderophile elements \citep{walker2009highly}. Large impacts are anticipated to be ubiquitous on young rocky planets (e.g., Mars: \citealt{nimmo2008implications, marinova2011geophysical, brasser2017colossal, marchi2020compositionally}, Venus: \citealt{gillmann2016effect, marchi2023long}), and while the impact scenarios presented in this study focus on an Earth-like planet, they demonstrate processes surrounding the accretion of impactor core material that are applicable to a wider range of planets in a similar mass range.
\begin{figure}
    \centering
    \includegraphics[width=0.99\textwidth]{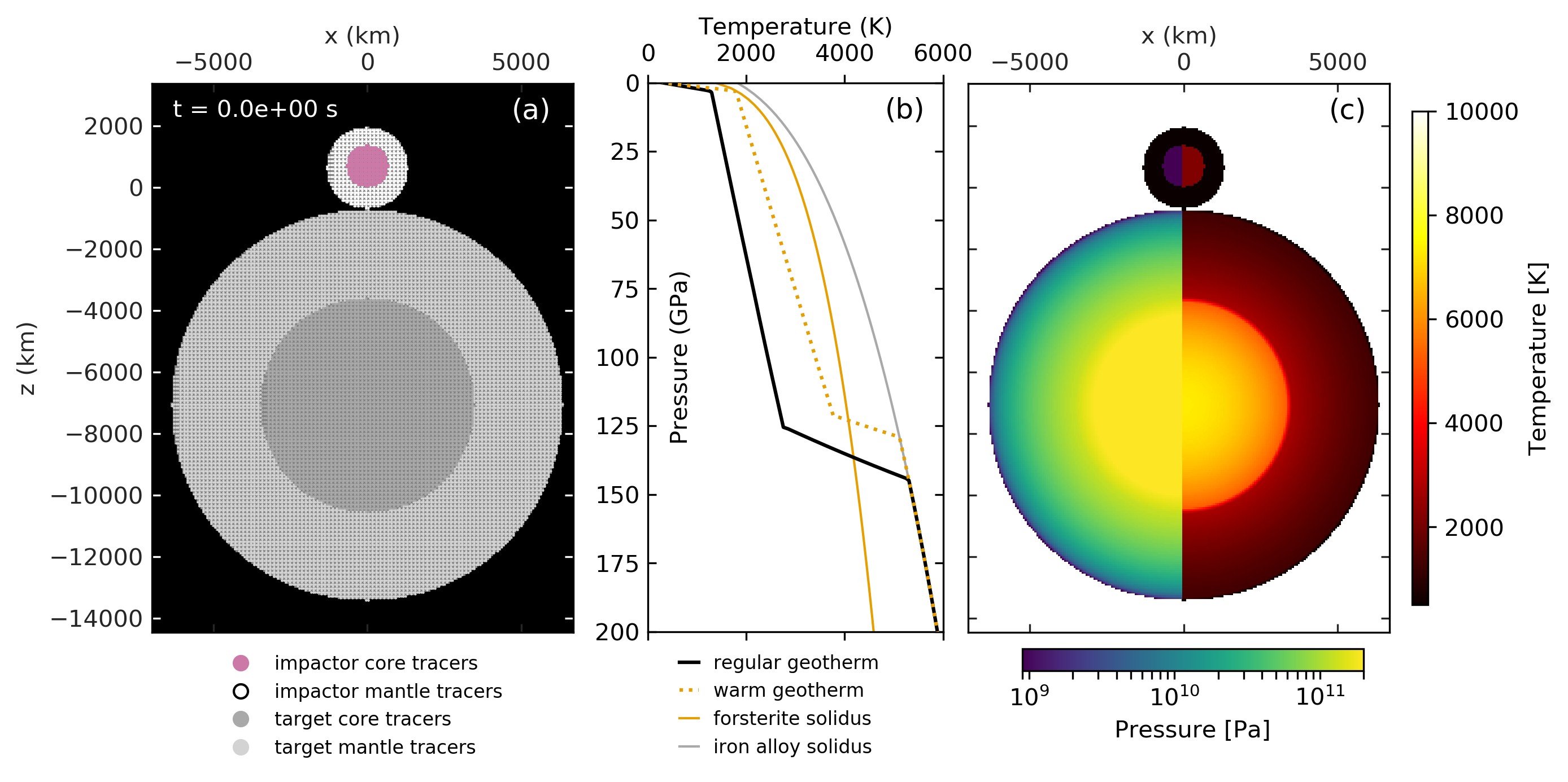}
    \caption{An Earth-like target and a $M_i = 3.604\times10^{22}\,\mathrm{kg}$ ($M_i/M_t = 0.0063$) impactor at the time of contact between the impactor and the target, with (a) the locations of tracer particles within the target and impactor, (b) geotherms under standard and warm mantle conditions (see Appendix \ref{sec:Setup Planet Init Temp}), as well as the solidus for our mantle and core materials, and (c) the pre-impact pressure and temperature profiles within the $y=0\,\mathrm{km}$ slice of both bodies.}
    \label{fig:Pre-Impact}
\end{figure}

Evidence for large impacts in Earth’s history after Moon formation ($4.5\sim4.3\,\mathrm{Ga}$) is visible in Earth’s mantle, where highly siderophile elements (HSEs) are present in excess of formation models’ predictions \citep{mann2012partitioning, rubie2015accretion}. These HSEs have strong tendencies to be incorporated into the core alongside iron during planet formation. Mantle excesses of HSEs, and particularly the elemental proportions in which we find them, thus indicate accretion of material with bulk-chondritic composition after core formation had ended \citep{rubie2015accretion, rubie2016highly}. Comparing Earth’s HSE excesses to those in the lunar mantle suggests that Earth accreted a greater total mass through late impacts than did the Moon, even accounting for their relative gravitational cross sections \citep{day2016highly}. Late delivery of $\sim 0.01 M_\Earth$ material to Earth with bulk chondritic composition is required to match the observed HSE abundances \citep{bottke2010stochastic, brasser2016late, day2016highly}. 

The power of large impacts to reduce planetary surfaces and deliver siderophile elements stems from the iron in the impactors' cores, which can react with the planetary atmosphere and interior upon break-up and accretion of the impactor material \citep{abe1988evolution, zahnle1988evolution, sleep1989annihilation, genda2017terrestrial, marchi2018heterogeneous, zahnle2020creation, citron2022large, itcovitz2022reduced, wogan2023origin}. How this metal is accreted to the planet is key in determining how reducing the post-impact surface environment can be and, conversely, how much reducing power is emplaced in other reservoirs during impact. \citet{genda2017terrestrial}, \citet{marchi2018heterogeneous}, and \citet{citron2022large} showed that upon break-up of the impactor, its core material is distributed between the target interior (as molten iron), the target atmosphere (as vapor or supercritical fluid), and escaping the target (gravitationally unbound from the planet). For smaller impact angles (i.e., closer to head-on collisions, $\theta_i = 0^\circ$) and lower impact velocities, a larger fraction of the metal is accreted by the target interior. Moving to greater impact angles (up to $\theta_i\sim 60^\circ$ from the vertical) and greater velocities, a greater fraction of the metal has sufficient kinetic energy to escape the target's gravitational field, and its reducing power is lost from the planet. 

\citet{itcovitz2022reduced} then showed that not only is the distribution of the impactor core into these three planetary reservoirs important, but that it also matters how the metal is accreted by the target interior. Due to interactions between the atmosphere and the impact-generated melt phase subsequent to impact, iron that is accreted by the mantle can still influence the redox state of the planetary atmosphere on short timescales. To achieve this influence, iron is required to reduce the melt phase, which then interacts with the atmosphere via gas exchange (i.e., melt should be mobile and able reach the surface of the melt pool via convection). The impactor iron's reducing power can thus be lost from the surface environment in one of three ways, by:
\begin{enumerate}[label={(\arabic*)}]
    \item{accreting to partially or entirely solid regions of the mantle (i.e., $< 40\%$ melt fractions), resulting in the iron's reducing power being locked away from the surface on solid, rather than melt, convective timescales ($> 100 \,\mathrm{Myr}$);}
    
    \item{accreting to the mantle but then sinking out and merging with the planet core;}
    
    \item{accreting to the mantle in sufficiently large fragments that the iron exists within the melt phase as large blobs, with large portions of the iron unable to chemically interact with the melt on its solidification timescale; this is in contrast to small droplets which are well-mixed with the melt and chemically available).}
    
\end{enumerate}
Phenomena (2) and (3) are somewhat related, with large blobs of iron being likely to sink out to the planet core before being able to break up into smaller chemically accessible units. \citet{itcovitz2022reduced} found that \ce{H2} abundances in the post-impact atmosphere can vary by up to an order of magnitude between the two end-member cases of all or none of the iron being available to reduce the melt. This is equivalent to several log units in oxygen fugacity ($f\ce{O2}$) for the melt-atmosphere system, and has consequences for both the ability of large impacts to produce globally reducing surface environments and the long-term redox state of the post-impact mantle. Note that, due to the oxidizing power of mantle material that is made accessible by melting (e.g., \citealt{schaefer2010chemistry, kuwahara2015molecular, marchi2016massive}), the redox power of the reducing impactor can be overshadowed to produce a net oxidizing effect of impact in some scenarios \citep{itcovitz2022reduced}. It is also important to note that these same arguments on the chemical availability of impactor core material are key in the deposition of HSEs into the planet mantle, which requires total equilibration between the host metal and the silicate melt to occur due to these elements' high metal-silicate partition coefficients \citep{kimura1992antarctic, borisov1997experimental, oneil1995experimental, jones2003signatures, mann2012partitioning}.

\citet{itcovitz2022reduced} based their calculations of iron distribution on the impact simulations outlined in \citet{citron2022large}. These smoothed-particle hydrodynamics (SPH) simulations do not incorporate material strength and are thus unable to fully capture the behavior of the impactor iron accreting to the target interior, as strength affects (i) the initial break-up of the impactor core, and (ii) whether impactor core material simply sinks to the planet core (strengthless) or is prevented from doing so by solid parts of the mantle that remain unmelted by the impact (requires strength). Here, we use an Eulerian shock physics code that includes material strength, iSALE, to tackle the end-member case problem set out in \citet{itcovitz2022reduced}. iSALE is well suited to simulations of \textit{large} impact events, as opposed to the \textit{giant} impacts to which SPH codes are commonly more suited (see Section \ref{sec:iSALE3D}).

In Section \ref{sec:iSALE3D}, we briefly describe iSALE, how we set up our models, and the parameter ranges that we examine in our suite of simulations. Section \ref{sec:Distribution} presents the distributions of impactor core material between different planetary reservoirs that we define. From these results, we then present a parameterization of the distributions as a function of a modified specific energy of impact. In Section \ref{sec:Discussion}, we compare our results and parameterizations to previous works, highlighting the effects of iSALE's ability to include material strength, and discuss the consequences of our results on the state of post-impact systems.

\section{iSALE3D: Setup and Parameters} \label{sec:iSALE3D}
The iSALE shock physics code \citep{wunnemann_strain-based_2006} is an extension of the SALE hydrocode \citep{amsden_sale_1980}. Modifications to the original SALE code include an elasto-plastic constitutive model, fragmentation models, various equations of state, and multiple materials \citep{melosh_dynamic_1992, ivanov_implementation_1997}. More recent improvements include a modified strength model \citep{collins_modeling_2004}, a porosity-compaction model \citep{wunnemann_strain-based_2006, collins_improvements_2011}, and a dilatancy model \citep{collins_numerical_2014}. iSALE, and its predecessors, have widely been used to determine melt volumes as a consequence of impact (e.g., \citealt{pierazzo1997reevaluation, pierazzo2000melt, wunnemann_numerical_2008, manske_impact_2021}). 
\begin{table}[t!]
    \centering
    \caption{Sizes and masses of bodies used in our main iSALE simulation suite. The grid cell size is set as $63.7\,\mathrm{km}$. A more extensive version of this table exists in Appendix \ref{sec:Setup Params}, covering additional test scenarios that were investigated.}
    \label{tab:Parameters}
    \begin{tabular}{rlllll}
        \toprule
        Parameter  &  Units   &  Target  &  Small Impactor  &  Medium Impactor  &  Large Impactor \\
        \midrule
        Mass   &  kg          &  $5.755\times10^{24}$  &  $2.775\times10^{21}$  &  $8.379\times10^{21}$  &  $ 3.604\times10^{22}$  \\
        \quad  &  $M_\Earth$  &  1.0  &  0.0005 &  0.0015  & 0.0063 \\
        \midrule
        No. of Cells (Mantle + Core)  &  -      &  200    &  18    &  26    &  42    \\
        Diameter (Mantle + Core)      &  km     &  12740  &  1147  &  1656  &  2675  \\
        \midrule
        No. of Cells (Core)           &  -      &  110    &  10    &  14    &  22   \\
        Diameter (Core)               &  km     &  7007   &  637   &  892   &  1401 \\
        \midrule
        Core-Mantle Mass Ratio        & -       &  0.366  &  0.259 &  0.255 &  0.244 \\ 
        \midrule
        Tracers in Impactor Core      & -       &  -      &  216   &  640   &  2640  \\
        \bottomrule
    \end{tabular}
\end{table}

In this study, we use the 3D version of the code, iSALE3D \citep{elbeshausen2009scaling}. The 2D code is limited to simulating head-on collisions due to its use of cylindrical symmetry, meaning that it cannot vary impact angle, a parameter which has been shown to be important in determining the break-up of the impactor during impact \citep{genda2017terrestrial, marchi2018heterogeneous, citron2022large}. To simulate the impact scenarios described in this work, several modifications to the iSALE3D code were required. A previous limitation of the code was that only one material interface was possible in each computational cell, meaning that scenarios with multiple differentiated objects were not possible. The code was recently modified to calculate multiple material interfaces per cell and update the material fluxes accordingly (\citealt{davison2022highMet, davison2022highLPSC}; Davison et al. in prep.). Additional improvements were made to the parallel performance of the code (Davison et al. in prep.), enabling the simulations described here to be run at their resolution in a much shorter time. One limitation of iSALE3D over iSALE2D is that it does not yet have the option to update the gravity field dynamically during the calculation. Instead, a so-called central gravity scheme is used that calculates the gravity field of the target planet based on its initial mass distribution, remaining fixed throughout the simulation. However, testing of comparable 2D and 3D central- and self-gravity simulations suggests that, for this regime of large impacts ($M_i/M_t \lesssim 0.01$), the difference between gravity schemes is negligible (Appendix \ref{sec:Setup Gravity Algorithms}).

\subsection{Materials and Thermodynamics} \label{sec:Materials and Thermodynamics}
The thermodynamic behaviors of materials are described by equations of state (EOSs) generated by the ANalytic Equation Of State program (ANEOS; \citealt{thompson_improvements_1974}). The mantle of each body is represented by an equation of state for forsterite \citep{collins2014improvements}, while the metallic core of each body is represented by an equation of state for iron (\ce{Fe_{85}Si_{15}}; \citealt{stewart2020equation}), both of which are tailored to the pressure and temperature regimes found in our large impact simulations (as opposed to giant impacts, where other EOSs are more suitable). The detailed properties of each material, including damage, strength and thermal softening model parameters, as well as melt curve Simon parameters, are detailed in Table \ref{tab:Material} in Appendix \ref{sec:Setup Params}. 

Pressure and temperature profiles are constructed within the bodies at each time step using the central gravity scheme and the material parameters. Earth's mantle is expected to have solidified from the Moon-forming impact during the time of Late Accretion (or indeed from any preceding impacts; e.g., \citealt{elkins2008linked, lebrun2013thermal, elkins2012magma, nikolaou2019factors, lichtenberg2021vertically}). We thus enforce an initial surface temperature of $293\,\mathrm{K}$, with a conductive thermal profile implemented for the thin lithosphere and a convective profile the planet mantle and core. Temperatures are limited to the material solidus (see \citealt{collins2016isale}) to ensure the correct rheology for the target (i.e., solid mantle, fluid strengthless core) in preparation for impact. Initial maximum temperatures of $\sim7500\,\mathrm{K}$ are found within the target core, and a pressure profile increasing $\sim 1\,\mathrm{GPa}$ to $\sim 400\,\mathrm{GPa}$ between the lithosphere and the center of the core (Figure \ref{fig:Pre-Impact}b,c). Our base-of-lithosphere temperature ($T_l = 1293\,\mathrm{K}$) is colder than expected for the Earth during the time of Late Accretion (e.g., \citealt{korenaga2008urey, davies2009effect, herzberg2010thermal, ganne2017primary, korenaga2021hadean}), resulting in our simulations providing a marginally conservative estimate of mantle melting and thus accretion to the planet core. However, we find that a geotherm towards the upper limit of expected mantle temperatures ($T_l = 1793\,\mathrm{K}$, Figure \ref{fig:Pre-Impact}b) only alters core accretion by $\lesssim 3\,\%$ (Appendix \ref{sec:Setup Planet Init Temp}). 

For all impacts, the vast majority of impactor mantle and core is melted and/or vaporized by the impact shock. Because of this, and because the temperatures and pressures of an impactor's post-shock state are orders of magnitudes greater than those found in its pre-shock state, each impactor's pre-impact thermodynamic state is inconsequential. Note that this also results in material strength in each impactor being inconsequential. As such, we assign a constant temperature profile for each of the mantle and core separately, using temperatures of $293\,\mathrm{K}$ and $2200\,\mathrm{K}$, respectively, as common sense values given the size of the impactors. Further to pressure and temperature, we track the Eulerian fields of density, material velocity, yield strength, and material volume fraction. These properties are updated at each time step of the simulation, with initial $\Delta t = 10^{-3}\,\mathrm{s}$ increasing to a maximum of $\Delta t = 10^{-1}\,\mathrm{s}$ as the impact progresses to represent the changing timescales of the dominant physics. Simulations are ran for a duration of $t = 3\,\mathrm{hours}$, at which point we find that motions have ceased and an approximate steady-state solution has been reached (see Sections \ref{sec:Setup Simulation Duration} \& \ref{sec:Setup Material Strength} for simulations ran with durations of $24\,\mathrm{hours}$).

In addition to our Eulerian grid values, we employ a Langrangian tracer particle system within the model \citep{pierazzo1997reevaluation}, enabling us to track the histories of material parcels as they move through the grid. Tracer particles are initially placed in the center of each grid cell within the target and impactor bodies (Figure \ref{fig:Pre-Impact}a) and then move time-dependently through the model using velocities based on the volumetric fluxes of their associated materials in and out of their current grid cells \citep{davison2016mesoscale}. Tracers effectively sample the Eulerian grid, and so also have associated temperatures, pressures, and velocities, as well as other tracked properties. We use the tracers, in tandem with the Eulerian material properties, to track how the impactor core material accretes to the target or escapes during impact (Section \ref{sec:Distribution}). 
\begin{figure}[p]
    \centering
    \includegraphics[width=0.88\textwidth]{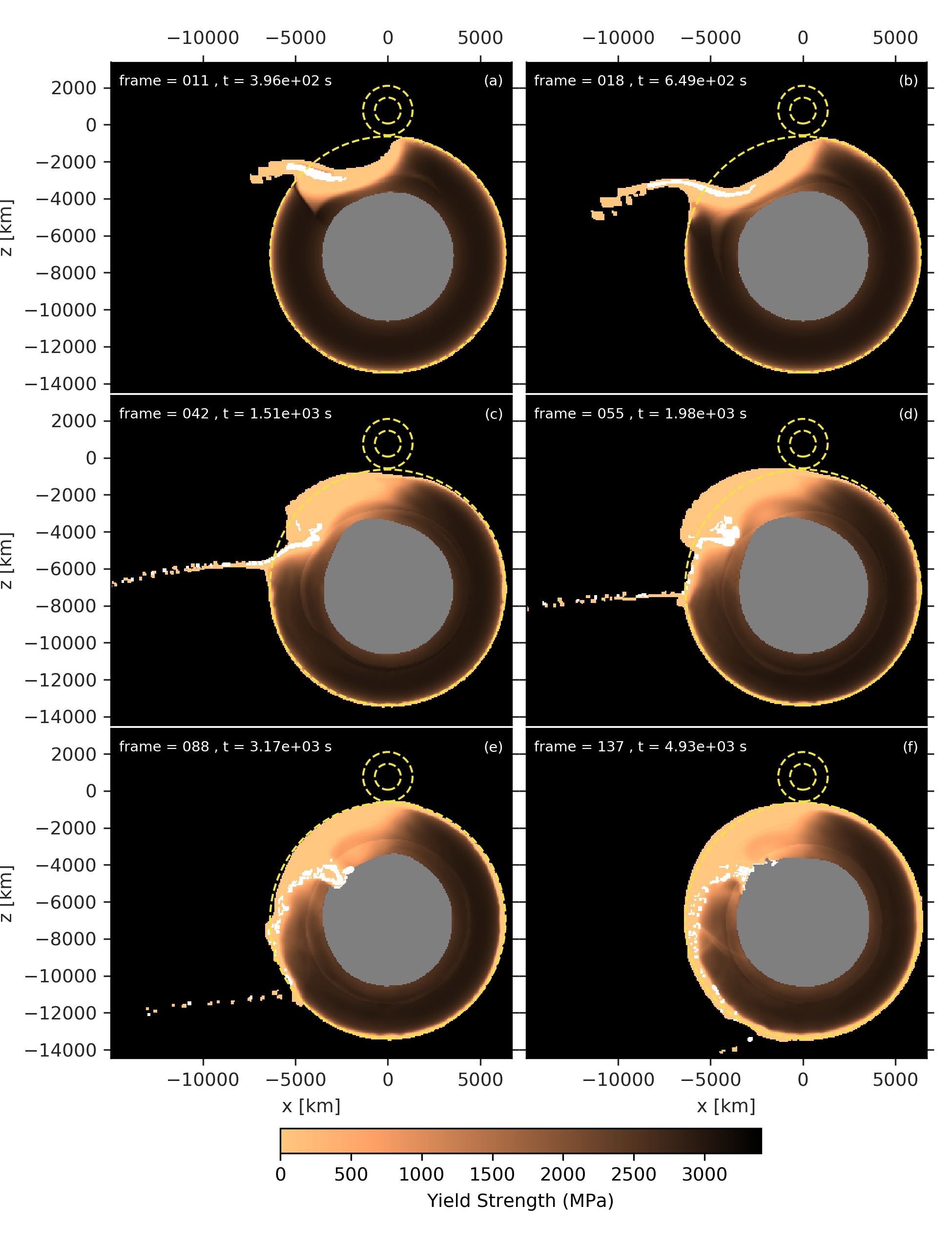}
    \caption{Evolution of the impactor core material, including (a,b) formation of a linear geometry of molten impactor iron through deceleration and pancaking, (c,d) disruption of the linear geometry, leading to material escape and formation of molten iron blobs and ligaments, (e,f) sinking of large molten iron blobs to the target core or residency in the mantle. Grid cells containing impactor iron are shown in white (cell mass fraction is ignored for simplicity but it should be noted that cells often contain $\ll 100\%$ iron). Yellow dashed line shows the initial location of the surface of the planet and the outer edges of the impactor mantle and core. Impact parameters: $M_i = 3.60\times10^{22}\,\mathrm{kg}$, $v_i = 1.5\,v_{esc}$, $\theta_i = 45^\circ$. The final time step for this impact (300, $t=3\,\mathrm{hours}$) is shown in Figure \ref{fig:Tracers}b.}
    \label{fig:Frames Iron}
\end{figure}

\vspace{-2mm}\subsection{Properties of the target planet and impactors} \label{sec:Target Impactor Properties}
We simulate three different impactor masses in our main simulation suite (Table \ref{tab:Parameters}). Our large-size impactor has mass $M_i = 3.66 \times 10^{22}\,\mathrm{kg}$, corresponding to approximately the mass of bulk chondritic material estimated as necessary to provide Earth's mantle HSE budget ($2.0\sim 4.8\times 10^{22}\,\mathrm{kg}$, \citealt{bottke2010stochastic, brasser2016late, day2016highly}). Our small-size impactor is ten times smaller than this with $M_i = 2.78 \times 10^{21}\,\mathrm{kg}$, which we chose due to impactors smaller than this not resulting in deep melting of the mantle, and as such being unlikely to result in impactor core material reaching the core. The mass of our medium-size impactor is in between the small and large masses at $M_i \sim 8.38 \times 10^{21}\,\mathrm{kg}$. Each impactor has an iron core approximately 0.3 of the impactor mass, although we underachieve this due to the restrictions placed by the grid resolution (Table \ref{tab:Parameters}). Additional impactor masses were simulated for specific investigations, with an $M_i \sim 5.10 \times 10^{21}\,\mathrm{kg}$ mass then incorporated into our main simulation suite (see Section \ref{sec:param_applicable}; Table \ref{tab:Parameters Extended} in Appendix \ref{sec:Setup Params}). An $M_i \sim 1.75 \times 10^{23}\,\mathrm{kg}$ was found to be too large to be captured in our accretion mode, and was not included (see Section \ref{sec:param_applicable}).

The target is Earth-like, with a mass of $M_t = 5.76 \times 10^{24} \,\mathrm{kg}$. This is slightly lower than Earth today, partially accounting for the yet to be accreted material of the Late Accretion impacts themselves, but also due to restrictions placed by the model resolution. The mass is constant between simulations so that initial conditions are identical. Variation in $M_t$ far beyond Earth-like values is beyond the scope of this work, but will be necessary for future analysis of these important planetary processing events, as the geometric and mass relations between the impacting bodies has an important influence on how impactor core material is accreted (Section \ref{sec:param_applicable}). Earth's crust is neglected due to its small size relative to the grid resolution. Any surface water or atmosphere present on the pre-impact target is also not resolvable in our simulations.

The standard grid cell size is set as $63.7\,\mathrm{km}$ such that the Earth-like target is 100 cells in radius. We additionally ran high and low resolution test simulations for comparative analysis using grid cell sizes of $36.4\,\mathrm{km}$ and $127.4\,\mathrm{km}$ respectively. The high resolution is approximately the greatest achievable resolution within the current implementation of iSALE3D. The low resolution doubles the size dimensions of the standard resolution we use. We find only small differences in the final distributions of impactor core material between the standard and high resolutions (Table \ref{tab:Results_Large_1}; Figure \ref{fig:Parameterization_Full_Detail} in Appendix \ref{sec:Results Table}). For a typical large impact ($M_i \sim 3.6\times10^{22}\,\mathrm{kg}$, $v_i \sim 1.5\,v_{esc}$, $\theta_i = 45^\circ$), there is an $\sim 0.08$ decrease in the mass fraction of material that ends up in the mantle from the standard resolution, corresponding to $\sim 0.04$ increases in both iron that ends up accreting to the target core and iron that is vaporized. The mass fraction of iron that escapes is approximately constant, although it is small in both cases. We suggest that these are expected behaviors that can be attributed to the higher resolution leading to shorter length scales within the materials, hence weakening structures within the bodies (i.e., a target mantle through which molten iron more easily sinks). Simulating at lower resolution demonstrates similar effects: we find greater mass fractions of the impactor core accreting to the mantle and less being vaporized or accreting to the target core than with the standard resolution (Table \ref{tab:Results_Large_1}). While we are restricted from increasing the resolution further, we propose that these simulations would converge in terms of the final iron distributions unless small iron droplets can be resolved, at which point simulation time would be excessive. We are thus confident that our standard simulation resolution of $63.7\,\mathrm{km}$ is an encouraging balance between computational cost and accurate capturing of material behaviors.

As well as varying impactor mass, our simulation suite examines the effect of impact angle and velocity on the distribution of iron. We carry out simulations at a range of angles between $0-70^\circ$, with the most frequent impact angle for an isotropic source of impactors being $45^\circ$ \citep{shoemaker1962interpretation}. Angles are defined relative to the vertical axis connecting the centers of the two bodies (i.e., $0^\circ$ is a head-on collision). We consider three impact velocities, defined as the relative velocity of the two bodies at the time of contact. Our slow impacts occur at the target-impactor mutual escape velocity ($v_{esc} \approx 11.2\,\mathrm{km\,s}^{-1}$ for our Earth-like target). Our medium velocity impacts occur at $16.0 \, \mathrm{km\,s}^{-1}$ ($\sim1.4 \, v_{esc}$) approximately the expected value for large impacts in the time after Moon formation \citep{raymond2013dynamical, brasser2020impact}. Our fast impacts have velocities of $20.0 \, \mathrm{km\,s}^{-1}$ ($\sim1.8 \, v_{esc}$). 

Compiled animations of selected simulations are available at \url{https://github.com/itcojo/impact_simulations}, showing the evolution of materials through time via the tracer particle system, as well as Eulerian grid values. 
\section{Distribution of Impactor Core Material} \label{sec:Distribution}
When an impactor penetrates into the Earth's mantle during a large collision, the impactor momentum and kinetic energy is partitioned between the impactor and target. This results in compression, heating, deceleration, and deformation of the impactor material. In oblique impacts, the trajectory of the impactor is also deflected upwards towards Earth's surface. The severity of these effects depends on the impact parameters ($M_i$, $\theta_i$, $v_i$), which in combination can be thought of in terms of the target mantle mass that the impactor intersects, and the momentum barrier that this mass represents. For example, more oblique or less massive impacts interact with a smaller fraction of the target mantle, resulting in less extreme material processing and hence a greater tendency to escape Earth's gravity well. The severity of these effects is deterministic in the fate of the impactor core material. As such, where the impactor iron ends up accreting to depends on both on the impact parameters and on the position of the material within the impactor core itself.

In all scenarios simulated here, within seconds of impact, the impactor core experiences extreme peak temperatures and pressures (thousands of Kelvin and $> 100 \,\mathrm{GPa}$). These are greater than the requirement for both incipient and complete melting of iron, meaning that all of the impactor core reach the material solidus (if not already molten before impact), while some of it will also reach the liquidus and become vapor or supercritical fluid (e.g., \citealt{kraus2015impact}). This is similarly true for the impactor mantle material, except large mass fractions can be vaporized due to the lower vaporization threshold of the silicates compared to the metallic core (e.g., \citealt{svetsov2005numerical}). The motion of the impactor material after the initial impact should, therefore, be thought of in terms of a hot viscous liquid. For the target mantle, deformation, melting, and vaporization are also substantial. The initial impact shock is responsible for some melting and much of the vaporization. However, a greater proportion of melt production is due to decompression melting and late-stage shear heating as the mantle adiabatically rebounds from excavation in the wake of the impactor's motion (Figure \ref{fig:Frames Iron}). Such rebounding motions occur on timescales on tens of minutes, after which the mantle settles out into a more spherical structure, with a large melt pool having formed downstream from the initial contact point.
\begin{figure}[t!]
    \centering
    \includegraphics[width=1.02\textwidth]{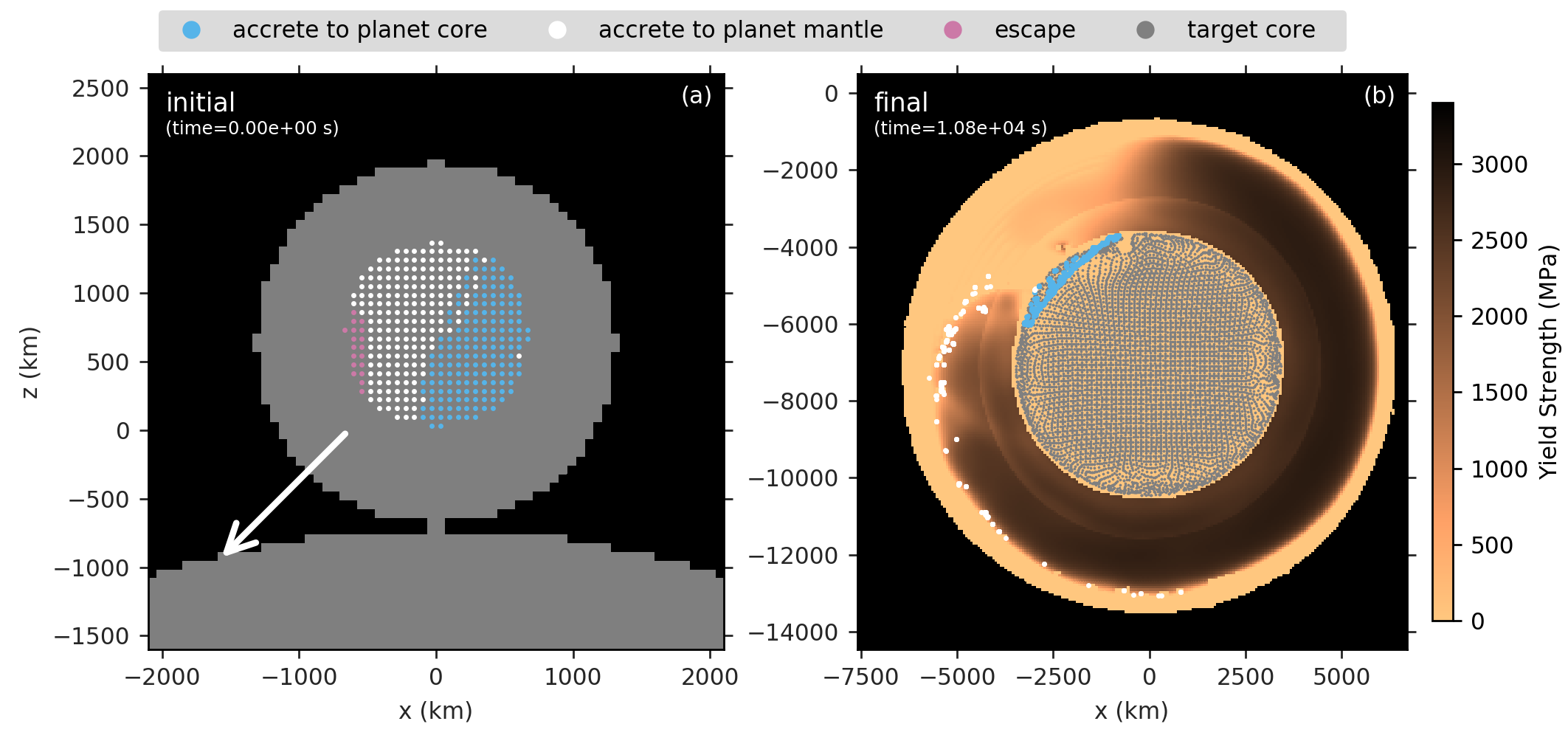}
    \caption{(a) Tracer particles of the impactor's metallic core (shown for the $y=0\,\mathrm{km}$ slice only) colored by the reservoirs that they are determined to accrete to at the end of the simulation; a white arrow is used to depict the impactor's velocity vector, and (b) locations of all impactor core tracer particles ($y>0\,\mathrm{km}$ also) at the end of the simulation ($t = 3\,\mathrm{hours}$), overlaid on top of the target mantle and core yield strengths ($0\,\mathrm{MPa}$ for fully fluid material). Impact parameters: $M_i = 3.60\times10^{22}\,\mathrm{kg}$, $v_i \sim 1.5\,v_{esc}$, $\theta_i = 45^\circ$.}
    \label{fig:Tracers}
\end{figure}

The different reservoirs to which impactor core material accretes are fundamental to any analysis of the post-impact system. In this Section, we detail how we classify impactor core material as accreting to a particular reservoir in our simulations, and present the distributions we find at the end of the simulations ($t = 3\,\mathrm{hours}$) as a function of impact parameters. We define three reservoirs: 
\begin{enumerate}
    \item{\textit{Escaped.} Impactor metal that dynamically escapes and is not accreted to the target on short timescales.} 
    \item{\textit{Accretes to target core.} Impactor metal that sinks to the planet core, the reducing power of which is lost from the melt-atmosphere system.}
    \item{\textit{Accretes to target mantle.} Impactor metal that is able to chemically affect the equilibrium state of the melt-atmosphere system, either through direct accretion to the impact-generated melt phase, or through vaporization during the initial impact, followed by condensation and rain-out to the molten surface.}
\end{enumerate}

\subsection{Escape} \label{sec:Escape}
We define escaping impactor core material as material that obtains sufficient kinetic energy to leave the target planet's gravity well in the immediate aftermath of the initial impact (i.e., timescales of hours). Such material may subsequently re-accrete to the target on million year timescales, or may be lost entirely from the target system through erosion and dispersion processes (e.g., \citealt{jackson2012debris}). Both of these fates are beyond the scope of this work, which focuses on shorter timescales.

We use the tracer particle system to track escaping impactor core material. At each model timestep, and for each tracer particle initially belonging to the impactor core, the velocity vector of the material to which the tracer particle belongs is compared to the escape velocity at the tracer's location:
\begin{equation}
    v_\mathrm{esc}(\mathbf{r}) = \sqrt{2|\mathbf{r}|~g(\mathbf{r})}\hspace{3mm},
    \label{eqn:Escape Velocity}
\end{equation}
where $\mathbf{r}$ is the positional vector between the target's center of mass and the considered tracer, and $g(\mathbf{r})$ is gravitational acceleration experienced at the tracer's location calculated under iSALE's central gravity scheme. If the velocity component of a tracer particle away from the planet's center of mass (i.e., $\mathbf{v}\cdot\mathbf{\hat{r}}$) reaches greater than $v_\mathrm{esc}(\mathbf{r})$, then the tracer, and the material it represents, is classified as having escaped the target.

We find that escape occurs predominantly for oblique impacts ($\theta_i \geq 45^\circ$), and for greater impact velocities and masses (Figure \ref{fig:Fraction Distribution}). For less oblique impacts ($\theta_i < 45^\circ$), impactor core material only escapes for the highest energy impacts in our simulation suite (large mass and fast velocity), and even then only a maximum of $\sim 1\,\%$ of iron mass escapes. The majority of simulations show no escaping core material. 

We find that iron from the impactor core escapes primarily from the leading edge of the impactor (i.e., the face of the impactor directed along its momentum vector; Figure \ref{fig:Tracers}). This is expected behavior, as although all of the impactor material begins each simulation at velocities greater than escape velocity, impactor core material away from the leading edge experiences substantial deceleration during impact due to the material and ejecta in front of it. A so-called pancaking of the impactor core thus occurs, whereby the spherical body spreads out into a flatter geometry in directions perpendicular to the initial momentum vector. Material at the leading edge of the impactor core hence generally experiences much smaller decelerations, and subsequently has a higher propensity to escape.

\subsection{Accretion to the Target Core} \label{sec:Core Accretion}
For a reasonably wide set of impact parameters, a melt column can form that penetrates down to the target planet's core-mantle boundary (CMB), enabling accretion of impactor core material to the planet core. We find that this column can either be generated directly by shock and decompression melting derived from the impact, or through the stresses acting on the mantle silicates caused by large clumps of sinking molten metal (i.e., the pressure caused by the weight of the metal leads to compressive stresses within the solid mantle material, leading to shear heating above the solidus and hence melting). This process also occurs in other mantle locations where impactor core material sinks down and meets solid mantle. However, the mass of metal interacting in these locations is generally much smaller than at the bottom of the melt column, leading to much shallower melt production in these locations.

The deceleration effects acting on impactor core material (Section \ref{sec:Escape}) can be extremely large, especially for impact angles $\lesssim 45^\circ$. Material can thus be decelerated such that its initial momentum is approximately entirely transferred to the planet, with the material's subsequent motion being dominated by the influences of gravity and the motion of the rebounding target mantle (Figure \ref{fig:Frames Iron}). Molten portions of the impactor core form blobs
\footnote{Blobs are approximately spherical, $m$- to $km$-scale, molten iron geometries that are commonly formed by the descending metal (e.g., \citealt{dahl2010turbulent}). Other geometries exist, such as sheets and ligaments \citep{villermaux2004ligament, shinjo2010simulation, villermaux2007fragmentation, kendall2016differentiated}, but are less common, and often break up into blobs either dynamically or through surface tension.} 
 within the target mantle, which we can resolve in our simulations down to the $63.7\,\mathrm{km}$ grid resolution. Even our smallest such blobs are still too large to be buoyantly suspended in the melt phase. Thus, molten metal in our simulations sinks through the impact-generated melt until it encounters either the target core or solid regions of the target mantle. In simulations without material strength, molten impactor core material sinks unencumbered through the fluid planet mantle and merge with the planet core (see Sections \ref{sec:Solid Mantle} \& \ref{sec:Setup Material Strength}).  

Physics that operates on length scales much shorter than the resolution of the simulations will lead to behaviors that cannot be captured in our simulations. Turbulent entrainment of silicates by the sinking metal blobs is one such physical phenomenon (e.g., \citealt{deguen2014turbulent, landeau2014experiments, landeau2021metal}), in which turbulence at the material interface drives convolution between the two phases. This entrainment of silicate material into the metal, and the potential liquid fragmentation of blobs into small droplets, will likely decrease in the amount of impactor core material that accretes to the planet core. However, quantification of this effect is not possible here.

We use the tracer particle system, in combination with the time-dependent volume fraction of each material in each grid cell, to calculate when metal from the impactor core reaches the target core. When we initialize the target and impactor bodies, we use two separate but identical materials (i.e., with the same material model; see Table \ref{tab:Material} in Appendix \ref{sec:Setup Params}) for the impactor and target cores, respectively. Thus, when a tracer particle from the impactor core reaches a cell containing target core material (threshold cell mass fraction $> 1\times 10^{-5}$), we register the impactor tracer as having reached this reservoir. This is only possible due to recent additions to iSALE3D, namely the multi-material capabilities (Section \ref{sec:iSALE3D}). This method could yield false positive accretion results if portions of the target core break off during impact and drift into the melt pool. However, we find that in almost all cases, the target core only deforms during impact, and portions do not separate off into the mantle, meaning that simulation cells containing target core material are exclusively part of the core itself. The only cases where this is not applicable are the large head-on collisions ($M_i \sim 3.60\times10^{22}\,\mathrm{kg}$, $\theta_i \lesssim 10^\circ$), where we do observe some small separation of material from the target core. However, the broken off portions of target core tend to sink back to the core by the end of the simulation, and these simulations do not form part of our parameterization (Section \ref{sec:param_applicable}).
\begin{figure}
    \centering
    \includegraphics[width=0.98\textwidth]{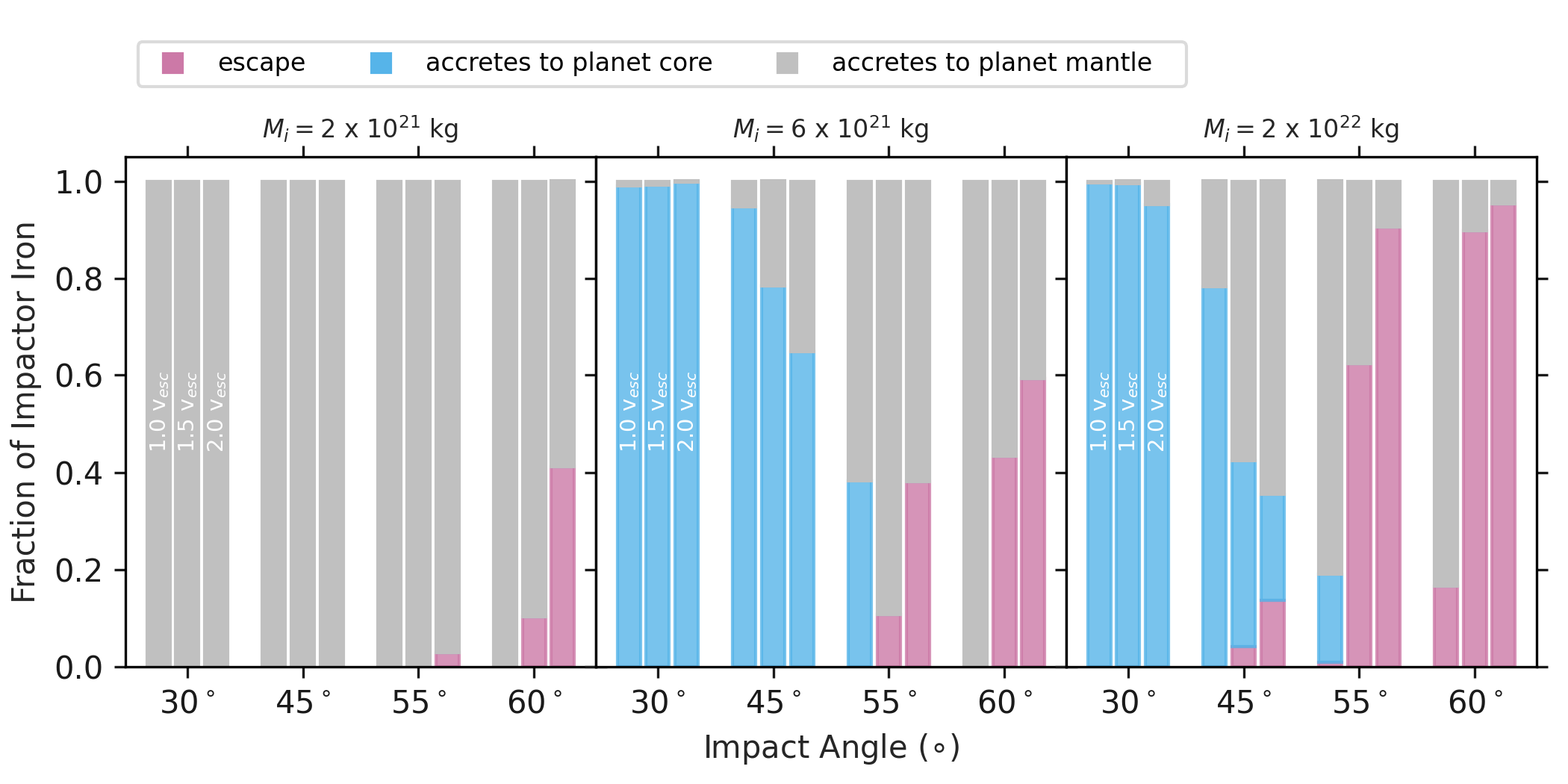}
    \caption{Mass fractions of impactor core material accreted to each of our defined accretion reservoirs for a selection of iSALE simulations, demonstrating the wide variety of distributions possible across the parameter space of large impacts. Numerical values are provided in Table \ref{tab:Results_Large_1} in Appendix \ref{tab:Results_Large_1}, alongside a much wider range of impact parameters.}
    \label{fig:Fraction Distribution}
    \vspace{2mm}
\end{figure}

We find that less oblique, larger, and slower collisions generally lead to greater accretion of the impactor core to the target core (Figure \ref{fig:Fraction Distribution}). Such impacts lead to a greater and/or more concentrated energy deposition in the location downstream of the impact site, and so are more likely to produce a melt column down to the CMB, a feature that is a necessary pre-requisite for iron accretion to the planet core. We find that accretion to the target core occurs primarily from the lower and rear portions of the impactor (Figure \ref{fig:Tracers}), reflecting both the material deceleration effects and geometrical effects relating to the location of the melt column.

\vspace{2mm}
\subsection{Accretion to the Target Mantle \& Vaporized Material} \label{sec:Mantle Accretion}
There are two different pathways for impactor core material to accrete to the target mantle. Firstly, impactor metal can accrete to the target mantle via the condensation and rainout of material vaporized during the initial stages of impact, leading to its relatively wide dispersal over the surface of the planet (e.g., \citealt{svetsov2005numerical, citron2022large}). Notably, metal accreted in this manner will exist as $mm$- to $cm$-scale droplets that can efficiently equilibrate with the molten mantle silicates \citep{rubie2003mechanisms, ulvrova2011compositional}, enabling the possibility of dissolution for sufficiently oxidizing melts near the planet surface.

Secondly, impactor core material can accrete directly as molten metal blobs that settle on top of regions of solid mantle. Such blobs can either be derived directly from the initial breaking apart of the impactor core (e.g., \citealt{genda2017terrestrial, marchi2018heterogeneous}), or can separate from larger iron blobs during the violent motion of the target mantle in the immediate aftermath of the impact (Figure \ref{fig:Frames Iron}). Molten iron settling out on top of solid mantle regions can descend through it on long timescales, either via formation of liquid metal diapirs or percolation of the iron through the stressed partially molten silicates (e.g., \citealt{zimmerman1999melt, samuel2012re, korenaga2023vestiges}), neither of which is accurately resolvable in iSALE. Such processes risk the impactor core material and its geochemical signature being lost from the mantle accretion reservoir to the planet core, and will thus be important for geochemical considerations on longer timescales after the impact (see Section \ref{sec:Distribution}). However, even in the unlikely end-member case where such processes are total, and all impactor core material that reaches the solid mantle eventually descends to the core, it is still important to distinguish between metal that has experienced this processing from metal that inherently merged with the core as a direct results of the impact itself.

In our simulations, we designate impactor core material that neither escapes nor accretes to the target core as accreting to the target mantle, accounting for both of the two mantle accretion pathways. We neglect the target atmosphere as an accretion reservoir, despite it being interpretable as a separate accretion reservoir for the vaporized iron material (e.g., \citealt{citron2022large}) as, for reasons of computational efficiency, the simulations described here do not track the evolution of material after it has been vaporized  (i.e., materials and tracer particles with a bulk density below $5\,\mathrm{kg ~m^{-3}}$ are deleted from model cells to increase calculation speed); our simulations are thus not well suited to giving an account of iron accreted to the target atmosphere. However, for the purposes of determining the state of the post-impact system, this should not be important. \citet{itcovitz2022reduced} showed that as long as the reducing power of the iron is available to either the atmosphere or the silicate melt, the equilibrium state of the interacting system will be the same. Combining together the two accretion pathways to the mantle is thus appropriate in this case. 

Material tends to accrete to the target mantle from the regions of the impactor core away from both the leading and rear edges of the core (Figure \ref{fig:Tracers}). Material from this region avoids the deceleration experienced by the impactor rear but still undergoes sufficient deceleration by material in front of it to slow to below escape velocity, and is thus dispersed downstream of where the melt column down to the planet core is located. Accordingly, we can observe a peak in the mass fraction of impactor core material accreting to the mantle as a function of impact angle. However, this optimal angle for accretion to the mantle varies with impactor mass due to variation in geometry (i.e., the mass of the target mantle that is intersected) and the resulting changes in the way kinetic energy and momentum are transferred during impact (i.e., compression, heating, deformation, and deceleration).
\section{Parameterization of Impactor Core Material Distributions}\label{sec:Parameterization}
The distribution of impactor core material during impact between the different planetary reservoirs defined in Section \ref{sec:Distribution} can be parameterized as a function of impact parameters, including the impactor mass ($M_i$), the impact velocity ($v_i$), and the impact angle relative to head-on ($\theta_i$). For each impact, the specific energy of impact is
\begin{equation}
    Q_R = \frac{1}{M_i + M_t} \left(\frac{1}{2} \mu v_i^2\right) \,\, ,
    \label{eqn:Impact Energy}
\end{equation}
\begin{equation*}
    \mu=\frac{M_i M_t}{M_i+M_t} \,\, ,
\end{equation*}
where $\mu$ is the reduced mass of the target-impactor pair. Modified forms of this specific impact energy account for the impact angle. For example, in the context of quantifying mantle melting during giant impacts, \citet{lock2017structure} proposed a modified specific energy of impact of the form,
\begin{equation}
    Q_S = Q_R \left(1 + \frac{M_i}{M_t}\right)(1 - \sin\theta_i) \,\, .
    \label{eqn:Modified Impact Energy}
\end{equation}
Empirical fits to our large impact simulations suggest a novel form of modified specific energy of impact, 
\begin{equation}
    Q_L = Q_R (1 - \sin\theta_i)^{-4} \,\, ,
    \label{eqn:Itcovitz Energy}
\end{equation}
which we find can be used to evaluate how impactor core material will be distributed between our defined accretion reservoirs. The physical interpretation of the functional form of $Q_L$ in relation to $\theta_i$ is not certain, despite empirically describing the simulation results well. However, we suggest that it likely captures the dependence of accretion on the volumetric fractions of interacting material between the impactor and the target \citep{leinhardt2011collisions} and/or the angular dependency of the momentum vectors along which the impactor material experiences deceleration.

The complexity of the physics operating in the simulations leads to numerous effects overlaying one another in space and time, and is ultimately why simulations are required. As described in Section \ref{sec:Distribution}, however, these effects can be viewed together as describing how the kinetic energy of the impactor is partitioned into the kinetic, internal energy and gravitational potential energy of the impactor and target materials during impact. The energy conversion is fundamental to the accretion process as it determines how the impactor core breaks up and the thermal and dynamic environments that this core material then experiences in the target mantle. 

\subsection{Accretion modes \& applicable impact domains}\label{sec:param_applicable}
\begin{figure}[t!]
    \centering
    \includegraphics[width=0.98\textwidth]{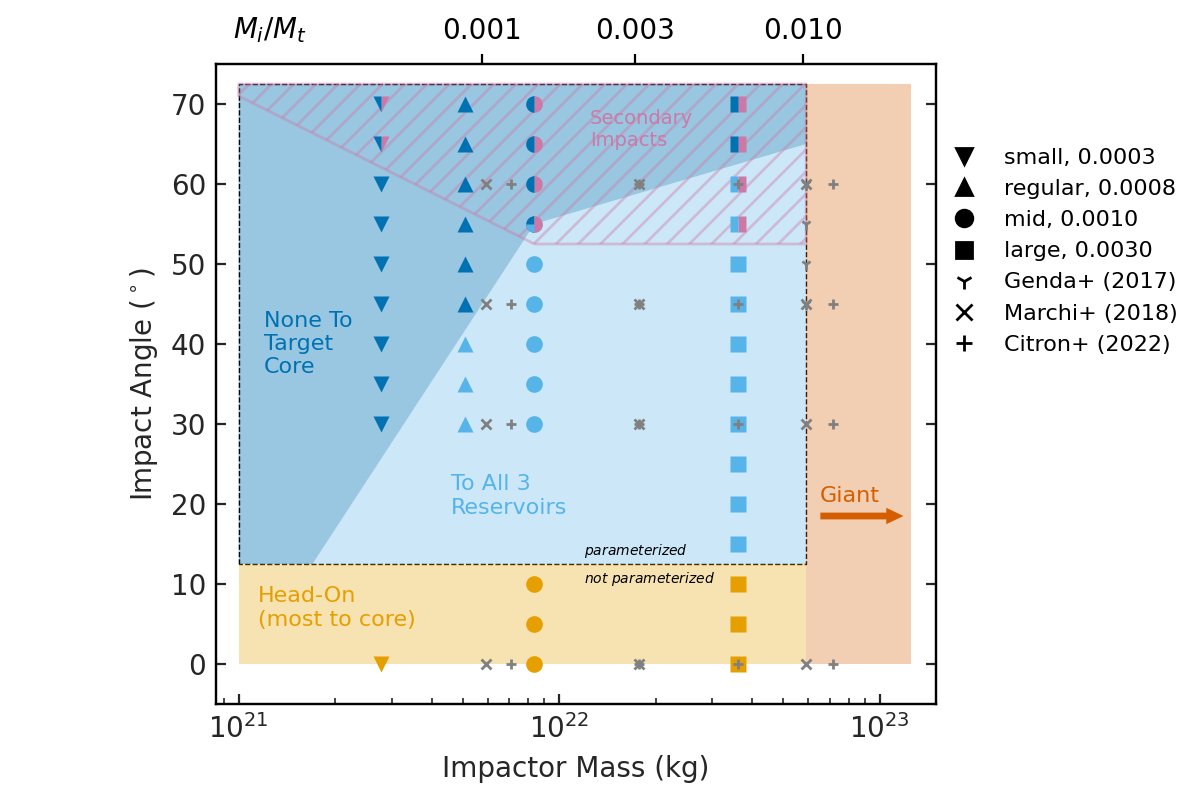}
    \caption{Regimes in the $M_i$-$\theta_i$ parameter space for which different accretion modes can be predicted to occur, where accretion modes correspond to the parameterizations we construct as a function of $Q_L$ (Section \ref{sec:param_fits}). Data points indicate iSALE simulations or SPH simulations of other studies. Data points with multiple colors indicate simulations at the same $M_i$-$\theta_i$ but with different $v_i$ that belong to different impact regimes. Note that regimes are not prescriptive, with both overlapping regions and boundaries between regions meant to indicate where the accretion of impactor core material is sensitive to impact parameters. In these regions, either of the numerical evaluations of Equation \ref{eqn:Parameter Fit} could be valid, or somewhere in between.}
    \label{fig:Fit Guide}
\end{figure}

We find three different modes of accretion in our simulations of large impacts: (1) approximately head-on impacts, (2) impacts that accrete to all three reservoirs of Section \ref{sec:Distribution}, and (3) impacts that are unable to accrete material to the target core (Figure \ref{fig:Fit Guide}). Each mode can be described by parameterizations as a function of $Q_L$ (Equation \ref{eqn:Parameter Fit}), with empirically derived limits on the impact parameter spaces for which they apply. Our parameterizations are described mathematically in Section \ref{sec:param_fits}. Here, we outline the parameter spaces in which each accretion mode operates and their associated parameterizations can be applied. There are potential extreme impacts that lie outside of these parameter spaces. However, the size-frequency distributions of impactors (e.g., \citealt{marchi2014widespread, brasser2016late}), and the probability distributions for impact angles \citep{shoemaker1962interpretation} and impact velocities \citep{raymond2013dynamical}, predict that the vast majority of large impactors with our bulk composition (i.e., core mass fraction $\sim 0.3$) that may have struck Earth in the period after Moon formation would fit within the criteria of our accretion modes.

Firstly, we find that each accretion mode operates up to a limit of impactor size. In terms of mass, we express this as requiring\footnote{Variations in $M_t$ for applications to larger planets (e.g., super-Earths) or smaller planets (e.g., Mars-like) are beyond the scope of this work. We thus do not include the $1 + M_i/M_t$ term used by \citet{lock2017structure} in our modified energy $Q_L$, despite this value being small for our impacts, as the results cannot necessarily be assumed to hold up to variations in $M_t$.} 
$M_i/M_t \le 0.01$. This can be described as the \textit{large} impacts domain\footnote{This could also be referred to as the \textit{sub-catastrophic} impacts domain to distinguish it from other impacts in Earth's history that have traditionally been referred to as \textit{large}, such as Chicxulub, Sudbury, or Vredefort. This difference in terminology is similar to the differences in how $\theta_i$ is defined, either from the vertical or from the horizontal, between research communities focused on planetary-scale impacts or asteroidal impacts, respectively. Here, we have elected to use \textit{large} in keeping with the terminology of previous related works.}, as opposed to the \textit{giant} impacts domain (i.e., $M_i/M_t > 0.01$) described by \citet{carter2020energy}, who considered how potential energy exchanges with kinetic and internal energies during such impacts. We performed simulations above this upper mass limit and indeed found that they sit outside the results space of the other simulations (see Tables \ref{tab:Parameters Extended} \& \ref{tab:Results_Large_1}, Figure \ref{fig:Parameterization_Full_Detail} in Appendix \ref{sec:Results Table}). Visual inspection of these simulations additionally shows that the usual processes of impactor core deceleration and break up do not occur in the same way for these giant impacts as with the large impactors, with a clearly diminished extent of impactor core pancaking, and hence a different morphology, at the end of the initial deceleration process. We do not present parameterizations for this giant impact accretion mode, as the lack of self-consistent gravity calculations within iSALE will likely begin to be an issue in this mass regime. However, we do present our results in Appendix \ref{sec:Results Table} for completeness. At the other end of the impactor mass scale, we find that there is no definitive lower bound as $f_{core} = f_{esc} = 0$ at sufficiently low impact energies, which is the same as if our parameterizations were not considered. Strictly, a lower bound may be set by requiring impactors to be differentiated (i.e., hundreds of kilometers; \citealt{neumann2012differentiation}), and should have a core-mantle mass ratio of approximately $0.3$ . 

\begin{figure}
    \centering
    \includegraphics[width=0.8\textwidth]{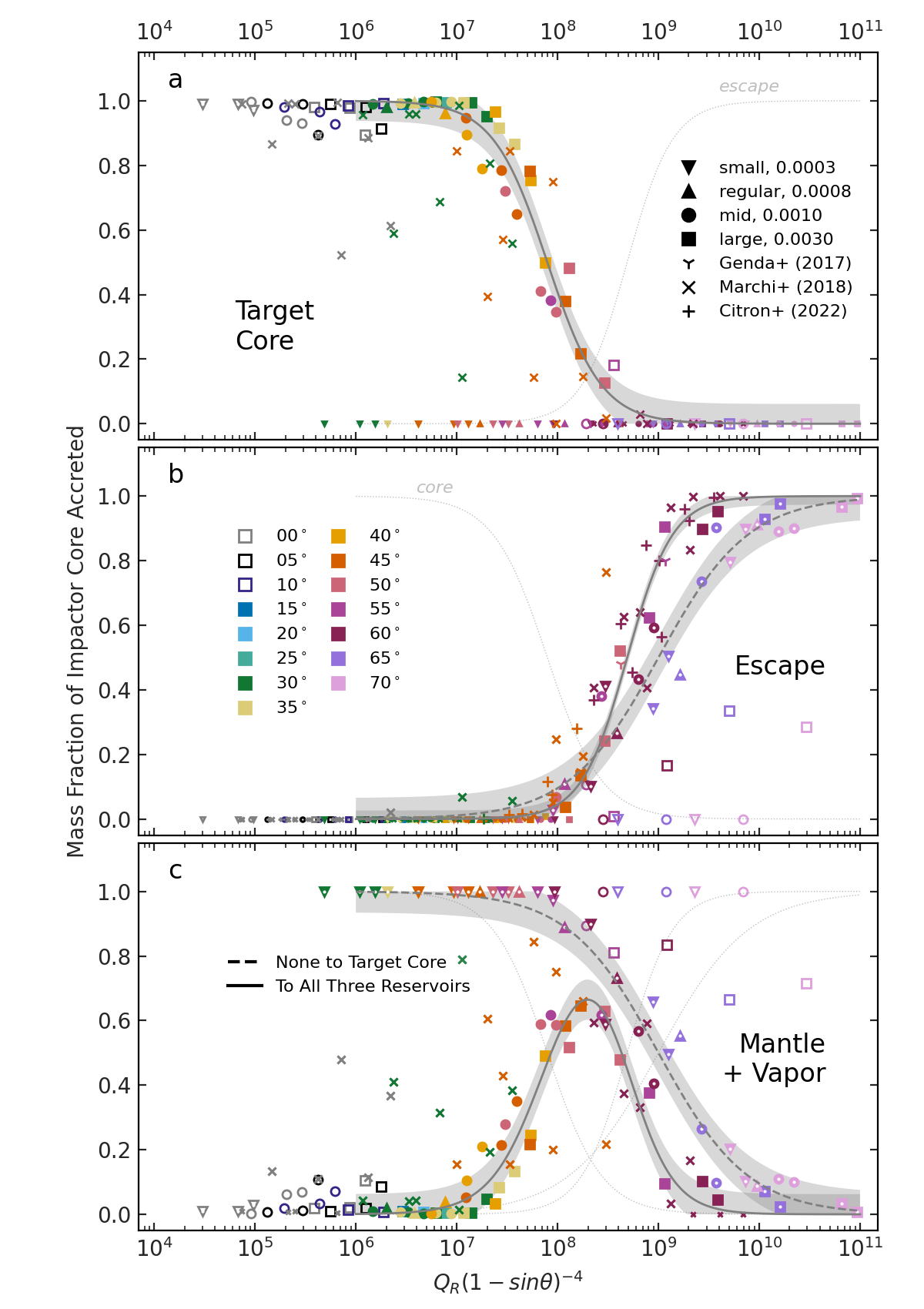}
    \caption{Mass fraction of the impactor core material, as a function of the modified specific energy of impact (Equation \ref{eqn:Itcovitz Energy}), that (a) accretes to the target core, (b) dynamically escapes and is not accreted to the target on short timescales, and (c) accretes to the target mantle, from iSALE simulations. Parameterized fits are shown by the solid and dashed lines, corresponding to different accretion modes, with a $1\sigma$ error contained by the shaded region around each line. Filled in data points are those included in the parameterized fits. Unfilled data points are those impacts which exhibit secondary impacts and are not included in the parameterized fits. Data from \citet{genda2017terrestrial}, \citet{marchi2018heterogeneous}, and \citet{citron2022large} is shown for comparison. Appendix \ref{sec:Results Table} contains an alternative version of this Figure with additional information on the impact parameters of \citet{marchi2018heterogeneous} and \citet{citron2022large}.}
    \label{fig:Parameterization}
\end{figure}

Variation in impact angle is another main source of difference in accretion mode. Impacts with $\theta < 10^\circ$ are approximately head-on impacts and are not well described by the parameterizations detailed in Section \ref{sec:param_fits}. The asymmetries in the excavation of the target mantle introduced by more oblique impacts lead to a rebounding motion of the mantle with a strong non-vertical component (e.g., Figure \ref{fig:Frames Iron}). Head-on collisions, on the other hand, mostly experience rebounding motions entirely back along the vertical axis. Conversion of impactor kinetic energy into potential and internal energy thus differs between the two scenarios, and the mantle environment (i.e., thermal structure and motion) experienced by the accreting core material is also consequently different, producing a different accretion mode. Our head-on collisions appear to follow a log-linear trend as a function of $Q_L$ (Figure \ref{fig:Parameterization}). However, the extreme environments generated in these impacts are not as well captured by our simulations as in the less extreme impacts at greater $\theta_i$ and so we do not present this log-linear trend here. Further, we expect these events to be uncommon in rocky planets' accretion histories, with only $\sim 3\%$ probability of occurring \citep{shoemaker1962interpretation}.

We find that there are regions of the velocity-angle parameter space where no accretion to the target core occurs. This is because accretion to the target core requires production of the melt column down to the planet CMB, a feature which impacts in this region of parameter space are unable to generate. Impactor core material thus only ends up either accreting to the target mantle or escaping. For smaller impacts, this parameter space is wide, with most impact angles resulting in no accretion to the target core (Figure \ref{fig:Fit Guide}). For larger impacts, this parameter space is restricted to highly oblique impacts only (e.g., $\theta_i \gtrsim 60^\circ$). A transition region is found in between, where small changes in impact angle or velocity can result in divergent accretion behaviors. The mass fraction of impactor core material undergoing escape in this accretion mode follows the same functional form as for the impacts that accrete to all three reservoirs, suggesting that they are closely related in their governing physics. However, they numerically have distinct parameters, and the two modes thus diverge with $Q_L$ (Figure \ref{fig:Parameterization}b). Importantly, the two modes strongly diverge in the mass fractions of impactor iron accreting to the target mantle due to the lack of inclusion of accretion to the target core (Figure \ref{fig:Parameterization}c). We emphasize that the boundary between these two modes is not prescriptive, especially as we have discretized impacts with $\Delta\theta_i = 5^\circ$ only. Rather, we intend that these boundary regions in parameter space be indicative of how impactor core material is likely distributed during impact. 

We find an anomaly when characterizing some impacts with high obliquity ($\theta_i \ge 55^\circ$). In cases of low impact velocities ($v_i \sim v_{esc}$), portions of impactor core material can be decelerated sufficiently that they no longer escape, but rather come back around on short timescales and undergo a secondary impact. The mass fraction of impactor iron that escapes is thus diminished and the mass observed as accreting to the target mantle is enhanced (Figure \ref{fig:Parameterization}, unfilled data points). The scale of the effect is unpredictable, with some secondary impacts leading to a difference in accreted iron of around $\sim30\%$, while others show a difference of $\sim60\%$. We thus do not classify these impacts as being a distinct accretion mode. Instead, we highlight the parameter space in which we find them (Figure \ref{fig:Fit Guide}), and advise care when using the parameterizations in this region. We exclude impact scenarios with secondary impacts from our fitting of the parameterizations.

\vspace{-3mm}\subsection{Numerical evaluation}\label{sec:param_fits}
We find that sigmoid functions describe well the mass fractions of impactor core material that accrete to each reservoir as function of $Q_L$. For both impacts that are able accrete to the target core and also those that cannot: 
\begin{equation}
    f = \frac{1}{1 + \exp[a\log_{10}Q_L + b]} \,\,\, .
    \label{eqn:Parameter Fit}
\end{equation}
For impacts that are predicted to be able to accrete material to the target core (Figure \ref{fig:Fit Guide}), $f_{core}$ can be calculated using $a_{core} = 3.78 \pm 0.28$ and $b_{core} = -7.90 \pm 0.02$ (RMSE $= 0.06$), and $f_{esc}$ can be calculated using $a_{esc} = -4.29 \pm 0.23$ and $b_{esc} = -8.67 \pm 0.02$ (RMSE $= 0.03$). For impacts that are predicted to be unable to accrete to the target core, only $f_{esc}$ is non-zero, and it can be calculated using $a_{esc} = -2.36\pm 0.18$ and $b_{esc} = -9.02 \pm 0.04$ (RMSE $= 0.06$). For each case, $|b|$ can be interpreted as the value of $\log_{10}Q_L$ that is the center of the sigmoid curve, and $|a|$ can be interpreted as the width of the sigmoid (i.e., the distance in units of $\log_{10}Q_L$ between $f=0$ and $f=1$). 

The mass fraction of impactor core material that is accreted to the target mantle, in accordance with how material accretion to this reservoir is defined (Section \ref{sec:Mantle Accretion}), is then given by:
\begin{equation}
    f_{mant} = 1 - f_{core} - f_{esc} \,\,\, ,
\end{equation}
where $f_{core} = 0$ in the appropriate parameter space and accretion mode.
\section{Discussion} \label{sec:Discussion}
Our suite of iSALE simulations represents impacts that are expected to have occurred during the period of Earth's accretion after the formation of the Moon. We have carried out over 130 simulations, covering a wide range of the \textit{large} impacts parameter space, including varying impactor masses, impact velocities, and impact angles, representing the first comprehensive study of large-scale planetary collisions using a grid-based shock physics code including material strength. From this data set, we have found that (1) there are different accretion modes within the \textit{large} impacts domain (Figure \ref{fig:Fit Guide}), (2) each mode has predictable trends in the mass fractions of impactor core material that are accreted to different reservoirs within and external to the target planet during impact (Figure \ref{fig:Parameterization}), and (3) these trends can be parameterized as a function of a modified specific energy of impact, $Q_L$. 

Here, we compare our parameterizations against previous studies based on the results of strengthless SPH simulations. We examine the differences in approaches and differences in physics between simulations, and explore how these differences affect the distribution of impactor core material during impact. We then briefly discuss the consequences of our findings for the geochemical evolution of young rocky planets and inferences about their early impact history, laying the foundations for future studies to use our body of work to inform more detailed geochemical calculations.

\subsection{Distributions of impactor core material}\label{sec:Comparisons}
Many of our impacts show small or zero escaping mass fractions of impactor core material (i.e., $f_{esc} \approx 0$). In fact, only highly oblique smaller impacts or moderately oblique larger impacts tend to show substantial $f_{esc}$. This is noticeably different to the accretion to the target core, for which most impacts with $M_i/M_t > 0.0010$ demonstrate some or considerable $f_{core}$. For much of the large impacts domain, therefore, impactor core material being sequestered to the planet core is the dominant source of loss of the metal from the planet mantle. 

Impacts modeled by \citet{genda2017terrestrial} (henceforth \citetalias{genda2017terrestrial}) all use $M_i/M_t = 0.01$ and $\theta_i \leq 60^\circ$, and should thus follow the accretion mode allowing for accretion to all three reservoirs (Figure \ref{fig:Fit Guide}), albeit sitting at the upper mass limit of this domain (Section \ref{sec:param_applicable}). The \citetalias{genda2017terrestrial} impacts predict similar $f_{esc}$ to our parameterization (mean error $\Delta f_{esc} = 0.01$, $\mathrm{RMSE} = 0.03$ , Figure \ref{fig:Parameterization}b). This is somewhat unexpected given the resolution of the \citetalias{genda2017terrestrial} models and the results of our resolution sensitivity tests: the $100\,\mathrm{km}$ size of their solid metal SPH particles is larger than our iSALE3D resolution ($63.7\,\mathrm{km}$), meaning that their particle smoothing length, $h$, the scale at which fluid behaviors are well treated by the SPH code, will be roughly double the size of our standard iSALE grid size. By halving the iSALE model resolution, we found that $f_{esc}$ was lowered substantially (Figure \ref{fig:Parameterization_Full_Detail}b, Appendix \ref{sec:Results Table}). The similarity of the \citetalias{genda2017terrestrial} data to our fits, with the caveat that there are only four data points, perhaps indicates that SPH codes and grid based hydrocodes are both suitable for analyzing the escape mass fractions in this impact regime.

SPH simulations performed by \citet{marchi2018heterogeneous} (henceforth \citetalias{marchi2018heterogeneous}) and \citet{citron2022large} (henceforth \citetalias{citron2022large}) have many similar parameters to one another, although it is important to note that they use separate implementations of SPH. These two data sets cover similar impact parameter spaces, mostly falling within the domain where accretion to all three reservoirs is possible (Figure \ref{fig:Fit Guide}). \citetalias{marchi2018heterogeneous} and \citetalias{citron2022large} use similar numbers of SPH particles, such that the numbers of particles in their impactor cores are similar to one another and to the number of tracer particles we have in our iSALE impactor cores; their particle smoothing lengths are both slightly longer than our grid resolution. We do not consider the largest impacts of either \citetalias{marchi2018heterogeneous} and \citetalias{citron2022large}, which are both above our $M_i/M_t = 0.01$ threshold, and as with our iSALE data, both data sets contain suggestions of secondary impacts taking place, with some low velocity impacts having substantially lower $f_{esc}$ and higher $f_{mant}$ than predicted by their $Q_L$ and accretion mode (Figure \ref{fig:Parameterization}).

We highlight these model similarities as, in addition to comparing the general trends of $f_{esc}$ and $f_{core}$ as a function of $Q_L$, they enable us to directly compare individual impacts between iSALE and each of \citetalias{marchi2018heterogeneous} and \citetalias{citron2022large}\,\footnote{We group both of the \citetalias{citron2022large} `escape' and `disk' reservoirs into a single reservoir, similar to how they are presented in the original work.}. \citetalias{marchi2018heterogeneous} consistently predict greater $f_{esc}$ than iSALE:  $\Delta f_{esc} = +0.09$, $\mathrm{RMSE} = 0.17$ for impacts accreting to all three reservoirs, and $\Delta f_{esc} = +0.01$, $\mathrm{RMSE} = 0.12$ for impacts unable to accrete to the planet core. \citetalias{citron2022large} also predict greater $f_{esc}$ but to a lesser extent: $\Delta f_{esc} = +0.05$, $\mathrm{RMSE} = 0.09$, and $\Delta f_{esc} = +0.02$, $\mathrm{RMSE} = 0.13$ for the same two respective reservoirs. The difference in $f_{esc}$ between iSALE and each SPH model increases as $v_i$ decreases. On the other hand,  \citetalias{marchi2018heterogeneous} consistently predict lower $f_{core}$ than iSALE ($\Delta f_{core} = -0.20$, $\mathrm{RMSE} = 0.32$), and the difference between the data sets now increases as $v_i$ increases. Investigation into the different EOSs used between simulations cannot rectify these discrepancies (Figure \ref{fig:Parameterization_Full_Detail}b, Appendix \ref{sec:Results Table}). Thus, considering all other similarities, we suggest that observed differences between the models are based in their differences in physics, namely material strength (Sections \ref{sec:Impactor Core} \& \ref{sec:Solid Mantle}).

For the accretion mode where accretion to all three reservoirs is possible, the two sigmoid curves for $f_{esc}$ and $f_{core}$ combine to create a peaked distribution for $f_{mant}$. There is thus an observable peak efficiency in accreting impactor core material to the target mantle, located at $Q_L \sim 2 \times 10^8 \,\mathrm{J \,kg^{-1}}$. In fact, $f_{mant}$ dominates over $f_{esc}$ and $f_{core}$ for the range $9 \times 10^7 \lesssim Q_L \lesssim 5 \times 10^8 \,\mathrm{J \,kg^{-1}}$. This region where $f_{mant}$ dominates is heavily populated by impact angles around $50^\circ$. However, we also observe some more oblique impacts in this region, tending to be smaller and faster impacts. There is thus a wide variety of impacts that can lead to deposition of substantial mass fractions of impactor core material into the target mantle, highlighting the need to account for a wide variety of impact parameters when considering the chemical and physical states of post-impact systems.

Mid-sized impactors ($M_i/M_t = 0.0015$) with $\theta_i = 55^\circ$ and large-sized impactors ($M_i/M_t = 0.0063$) with $\theta_i = 50^\circ$ offer interesting test cases; both sets of impacts (i.e., the three different velocities) straddle the peak in the $f_{mant}$ curve, highlighting some of the model nuances described both in this Section and in Section \ref{sec:Parameterization}. For the mid-sized impactors, the $1.0\,v_{esc}$ impact is able to produce melt down to the CMB, and hence shows non-zero accretion of impactor core material to the target core. The $1.5\,v_{esc}$, on the other hand, sits well above the peaked distribution at $f_{mant} \sim 0.9$, displaying secondary impact behavior. Lastly, the $2.0\,v_{esc}$ impact only undergoes escape (despite sitting below the curve fit for this accretion mode). These mid-sized impactors thus highlight the large differences that can be achieved in the distribution of impactor core material by making small changes in impact parameters near the boundaries of accretion modes. For the large-sized impactors, a completely different story is told, with all three velocities lying comfortably on the peaked $f_{mant}$ curve. These large-sized impactors thus not only emphasize the robustness of this accretion mode in this region of parameter space, but also demonstrate that the interesting underlying physics dictating the balance in distribution of core material between our three accretion reservoirs. 

\subsection{Break up of the impactor core}\label{sec:Impactor Core}
The initial phase of the impact ($\sim$ hundreds of seconds), in which deceleration and break up of the impactor core occurs, sets the morphology of the impactor metal from which all subsequent behavior follows. For escaping material, there are two relevant effects that may cause the differences between our iSALE simulations and the fluid SPH implementations of \citetalias{marchi2018heterogeneous} and \citetalias{citron2022large}. 

The first effect is that of the gravity algorithm used, in which we are restricted to the fixed central gravity regime. We ran investigative simulations (Appendix \ref{sec:Setup Gravity Algorithms}) that demonstrate that there is insubstantial difference between self-consistent and central gravity algorithms for impactor core material accreting to the planet interior. However, we are unable to draw such conclusions for material escaping due to the restriction of iSALE2D to head-on collisions and thus the lack of any escaping material (Appendix \ref{sec:Setup Gravity Algorithms}). We suggest, however, that greater values of $f_{esc}$, as seen in \citetalias{marchi2018heterogeneous} and \citetalias{citron2022large}, could plausibly arise from the diminished gravity field of a self-consistent calculation, with this difference being caused by the impact-derived cavity in the planet mantle.

The second potential cause of lower $f_{esc}$ in our iSALE simulations is the inclusion of material strength, damage models, and plasticity models in the planet's mantle. Impactor core material undergoing escape tends to come primarily from the leading edge of the impactor (Section \ref{sec:Escape}, Figure \ref{fig:Tracers}), a region that experiences diminished deceleration in comparison to the material behind it and dynamically breaks off during the pancaking process (Figure \ref{fig:Frames Iron}a,b,c). Under iSALE's material models, deformation of the mantle requires greater energy than if it were modeled as an elastic (or viscoelastic) fluid, with the kinetic energy of the impactor being the source of this additional energy. We would thus expect a diminished portion of this leading edge to retain velocities exceeding $v_{esc}$ than in a purely fluid system. Indeed, in a test simulation that we ran with and without material strength (similar to that shown in Appendix \ref{sec:Setup Material Strength} but now with $\theta_i = 55^\circ$ in order to obtain a greater $f_{esc}$), we found an increase from $f_{esc} = 0.624$ with strength to $f_{esc} = 0.656$ in a fluid simulation without strength.

Both effects can explain why we see increasing differences in $f_{esc}$ between iSALE and the two SPH models as impact velocity decreases. Faster impacts are more likely to have sufficient energy to overcome the additional energy barriers presented by either the additional gravitation pull of central gravity or the additional deceleration caused by the material properties. Slower impacts are less likely to have sufficient energy and so will be more sensitive to these effects. Lastly, this applies to slow impacts that generate secondary impacts, for which a greater mass fraction of the impactor core is retained (i.e., lower $f_{esc}$) in these sub-escape velocity clumps in iSALE than in the SPH simulations. 

\subsection{Treatment of solid mantle regions}\label{sec:Solid Mantle}
The inclusion of material strength in iSALE enables our simulations to capture the interactions of fluids with solids. This importantly includes encounters between sinking impactor core material and solid regions of the mantle that remain unmelted during the impact. As a result, metal accreting to the planet interior can come to rest on top of solid mantle regions (Figures \ref{fig:Frames Iron} \& \ref{fig:Tracers}), even if a melt column down the planet's CMB exists at other azimuthal locations. Iron accreting in this manner remains trapped in the mantle during our simulations and contributes substantially to our $f_{mant}$. Such behavior is markedly different from strengthless simulations, where all materials are treated as fluids and molten metal sinks through the silicate mantle to the planet core in all cases, purely through density differences between the two materials (see Appendix \ref{sec:Setup Material Strength} for an example of a strengthless iSALE simulations where such sinking occurs, demonstrating the importance of the feature within the model). 

We might thus expect $f_{core}$ to be enhanced in the \citetalias{marchi2018heterogeneous} simulations when compared to equivalent impacts in iSALE. However, this is not what we observe in the data, with \citetalias{marchi2018heterogeneous} predicting lower $f_{core}$ than our results on an impact-by-impact basis (mean error of $\Delta f_{core} = 0.20$). There are a number of compounding effects that could be responsible for this. One example is the lower resolution of \citetalias{marchi2018heterogeneous}'s particle smoothing length when compared to our grid size, which our sensitivity tests indicate decreases $f_{core}$ (Figure \ref{fig:Parameterization_Full_Detail}a, Appendix \ref{sec:Results Table}). However, the difference seen between the two data sets is much greater than the difference indicated by this resolution test. Alternatively, the purely fluid nature of the planetary interior in \citetalias{marchi2018heterogeneous} may result in impactor iron being suspended by long-lasting oscillations in the planet mantle (e.g., \citealt{emsenhuber2018sph}), which are prevented in the iSALE simulations by the presence of the solid material parameters (Appendix \ref{sec:Setup Material Strength}). If so, when such motions cease (which occurs on timescales longer than the total simulation time of \citetalias{marchi2018heterogeneous}), we would expect the molten iron to descend to the planet core. 

However, there is generally limited agreement between the data sets (Figure \ref{fig:Parameterization}a), with clear and obvious differences in the shapes and spreads of the data sets, and sometimes vast differences in $f_{core}$ when comparing similar impacts between the two data sets. We thus suggest that the fundamental differences in the treatments of solid regions of the mantle mean that the two sets of simulations simply represent two different accretion processes: one process requiring a melt column down the CMB for accretion to the planet core, and one where iron may sink to the planet core at any azimuthal position. The presence of material strength, and the associated solid-fluid interactions, in iSALE should yield a more accurate characterization of impactor core material's behavior during accretion to the planet interior than in purely fluid models (even those that artificially enhance the viscosity of mantle determined to be solid by thermodynamic prescriptions, which \citetalias{marchi2018heterogeneous} do not). However, the necessity of generating a melt column down to the CMB in order for accretion to the planet core to occur does highlight the importance of accurate characterization of the extent and geometry of mantle melting during planetary-scale impacts. While our simulations are well suited to such calculations, future changes in physics or prescriptions within iSALE or other impact simulations that include material strength may provide additional insight into this phenomenon.

\subsection{Geochemistry \& Post-Impact Consequences}\label{sec:Consequences}
Studies considering the consequences of large impacts on young rocky planets must commonly invoke many assumptions, such as the pre-impact state of the impacted planet or the accretion patterns of the impactor core material. Our extensive simulation suite means that accurate predictions can now be made for the fate of the impactor core, which can be taken forwards into calculations on the chemistry of the atmosphere and interior in the short- and long-term after impact. Here, we briefly demonstrate some such uses of our findings.

The oxygen fugacity of a post-impact atmosphere can vary by up to two orders of magnitude (or equivalently an order of magnitude difference in \ce{H2} abundance; \citealt{itcovitz2022reduced}) between $0-100\%$ of impactor iron that is accreted to the planet interior being chemically available to the melt-atmosphere system. \citet{itcovitz2022reduced} speculated that there would be low iron availability if substantial fractions of the impactor core accreted as large blobs that efficiently sink to the planet core. Our simulations show that this is indeed possible for impacts with $M_i/M_t \gtrsim 0.0009$ and $\theta_i \lesssim 50^\circ$. Indeed, we find $f_{core} > f_{mant}$ for many impacts up to $\theta_i \sim 45^\circ$, showing that less reducing post-impact environments may be common. Depending on the impactor mass and the initial oxygen fugacity of the planet mantle, large impacts may even lead to a net oxidization of the environment \citep{marchi2016massive, itcovitz2022reduced}.

However, we have also demonstrated that the production of a large impact-generated melt pool does not necessitate high $f_{core}$ and the locking away of the impactor's reducing power from the surface. $100\%$ availability of impactor iron to the mantle and surface is only possible for the smallest of our impactors, while we find an upper limit of $\sim 65\%$ availability for larger impactors. Nonetheless, given the high likelihood of occurrence for impact parameters that lead to greater accretion to the planet mantle ($\theta_i \sim 45^\circ$, \citealt{shoemaker1962interpretation}; $v_i \sim 1.5~v_{esc}$, \citealt{raymond2013dynamical}), large differentiated impactors commonly accrete greater than $40\%$ of their metallic cores to the planet mantle. The ability of large impacts to produce reducing environments in such cases then depends on the efficiency of the accreted metal in equilibrating with the melt phase.

The traditional analytical approach to this problem involves a disruptive cascade of large blobs down into small blobs and then droplets (e.g., \citealt{rubie2003mechanisms, ichikawa2010direct, qaddah2019dynamics, maas2021fate}). Experiments on the fragmentation of liquid volumes, however, suggest that silicates of the mantle melt phase mix with the descending molten impactor metal through turbulent entrainment, with the potential for liquid fragmentation of blobs into droplets directly if a sufficient silicate mass is entrained (e.g., \citealt{deguen2014turbulent, landeau2014experiments, landeau2021metal}). Recent modeling suggests that rapid three-phase flow with solid silicate, molten silicate, and molten impactor metal could lead to retention of the impactor core once it sinks to the bottom of the impact-generated melt phase \citep{korenaga2023vestiges}. There are thus several ways to enable equilibration between impactor core material that accretes to the planet mantle and the mantle itself. However, such methods are not complete, certainly on short timescales, and there is still the risk of losing molten metal to the core from the top of the solid mantle.

Geochemical evidence in Earth's mantle for large impacts in the time after Moon formation, including HSE abundances and isotopic signatures, is similarly influenced by the impactor iron distribution. If we assume that at least one of the proposed equilibration mechanisms between the impactor core material and the planet mantle is efficient, our simulations demonstrate that a variety of impacts can generate notable metal-derived geochemical fingerprints in the planet mantle. Indeed, combinations of impact parameters (mass, velocity, angle) that may have been expected to be inefficient at depositing impactor core material into the planet mantle from previous simulations (e.g., $M_i = 8.4\times 10^{21}\,\mathrm{kg} = 0.0015\,M_\Earth$, $\theta_i = 55^\circ$) have been shown to be among the most efficient (Figure \ref{fig:Parameterization}). We have also shown that many impacts, particularly those with low or high obliquity, will leave almost no geochemical signal reflecting impactor core material in the mantle, with the metal instead escaping or being sequestered to the planet core. Such impacts thus may have occurred but have left little trace in Earth's present day geochemistry.

Previous calculations of Earth's accretion history (e.g., \citealt{bottke2010stochastic, brasser2016late}) have commonly neglected such considerations, assuming that $100\%$ (or close to; \citealt{raymond2013dynamical}) of impactor core material contributes to the HSE or isotopic signature of an impact, although several calculations have taken the possibility of such effects into account (e.g., \citealt{rudge2010broad, jacobson2014highly}). Studies that neglect this inefficiency in the recording of each impact will thus have likely underestimated the bombardment flux of impactors onto early Earth. \citet{genda2017terrestrial}, \citet{marchi2018heterogeneous}, and \citet{citron2022large} made steps into removing the need for assuming accretion efficiencies. Here, we have greatly expanded the explored parameter space, allowing for more accurate characterization of the chemical consequences of large impacts across a majority of possible scenarios, and enabling the construction of more complex and nuanced accretion histories for Earth by future works. 
\section{Conclusions} \label{sec:Conclusion}
For many aspects of the early evolution of rocky planets, large impacts can play a fundamental role, transforming their chemical environments and reshaping the atmospheric composition. Key to these processes is the availability of an impactor's core material to drive reduction at a planet's surface and deliver metals to its mantle. We have presented 3D simulations of such impacts using the shock physics code iSALE to determine the fate of the impactor core. For each impact scenario, we have calculated the mass fractions of this core material that escapes the planet ($f_{esc}$), sequesters to the planet core ($f_{core}$), or accretes to the planet mantle ($f_{mant}$). Further to this, we have constructed parameterizations for the distribution of impactor core material between these reservoirs as a function of impact parameters, defining two possible accretion modes ($f_{core} = 0$ and $f_{core} \neq 0$). 

Our work demonstrates that the inclusion of material strength in iSALE enables more accurate characterization of sequestration of impactor core material to the target core than in strengthless models. This is achieved through enabling the settling out of molten iron onto solid unmelted parts of the target mantle. The inclusion of material strength thus leads to improved evaluations of $f_{core}$ in comparison to those found by SPH codes, for which the implementation of accurate strength models is inherently challenging in a way that it is not for grid-based Eulerian codes. We additionally find that material strength may contribute towards the early dynamic behavior of impactor material, which indirectly affects the mass fraction of the impactor core that escapes. However, this effect is small and may alternatively be caused by our current inability to calculate perturbations to the planet's gravity field caused by excavation of the mantle during impact. 

For a considerable subset of impact parameters, we find that sequestration to the planet core represents a greater source of lost reducing power than escape. However, for likely impact parameters, we find that $f_{mant} > 40\,\%$ is common, even for impacts that have previously been considered to not contribute significantly to mantle-accreted material (e.g., $\theta_i = 55^\circ$). Impactor core material that remains in the mantle may be chemically available to drive reduction of the post-impact atmosphere and deposit geochemical evidence of the impact's occurrence (e.g., highly-siderophile elements or isotopic signatures) if mechanisms for equilibration between the metal and the silicate melt phase are efficient. If so, given the ubiquity of planetesimal-sized debris following planet formation and protoplanetary disc dissipation, it may be common for large impacts to generate reducing atmospheres on young rocky worlds. However, substantial mass fractions of impactor core material can yet be lost from the planet mantle, leading to underestimates in the inferred total mass of material accreted to Earth during the period of Late Accretion based on present day geochemical evidence.

\bibliographystyle{aasjournal}


\begin{thebibliography}{}
\expandafter\ifx\csname natexlab\endcsname\relax\def\natexlab#1{#1}\fi
\providecommand{\url}[1]{\href{#1}{#1}}
\providecommand{\dodoi}[1]{doi:~\href{http://doi.org/#1}{\nolinkurl{#1}}}
\providecommand{\doeprint}[1]{\href{http://ascl.net/#1}{\nolinkurl{http://ascl.net/#1}}}
\providecommand{\doarXiv}[1]{\href{https://arxiv.org/abs/#1}{\nolinkurl{https://arxiv.org/abs/#1}}}

\bibitem[{Abe \& Matsui(1988)}]{abe1988evolution}
Abe, Y., \& Matsui, T. 1988, Journal of the Atmospheric Sciences, 45, 3081

\bibitem[{Amsden {et~al.}(1980)Amsden, Ruppel, \& Hirt}]{amsden_sale_1980}
Amsden, A.~A., Ruppel, H.~M., \& Hirt, C.~W. 1980, {SALE}: {A} simplified {ALE} computer program for fluid flow at all speeds, Tech. Rep. LA-8095, Los Alamos Scientific Lab., NM (USA), \dodoi{10.2172/5176006}

\bibitem[{Benner {et~al.}(2020)Benner, Bell, Biondi, Brasser, Carell, Kim, Mojzsis, Omran, Pasek, \& Trail}]{benner2020when}
Benner, S.~A., Bell, E.~A., Biondi, E., {et~al.} 2020, ChemSystemsChem, 2

\bibitem[{Borisov \& Palme(1997)}]{borisov1997experimental}
Borisov, A., \& Palme, H. 1997, Geochimica et Cosmochimica Acta, 61, 4349

\bibitem[{Bottke {et~al.}(2010)Bottke, Walker, Day, Nesvorny, \& Elkins-Tanton}]{bottke2010stochastic}
Bottke, W.~F., Walker, R.~J., Day, J.~M., Nesvorny, D., \& Elkins-Tanton, L. 2010, science, 330, 1527

\bibitem[{Brasser \& Mojzsis(2017)}]{brasser2017colossal}
Brasser, R., \& Mojzsis, S. 2017, Geophysical Research Letters, 44, 5978

\bibitem[{Brasser {et~al.}(2016)Brasser, Mojzsis, Werner, Matsumura, \& Ida}]{brasser2016late}
Brasser, R., Mojzsis, S., Werner, S., Matsumura, S., \& Ida, S. 2016, Earth and Planetary Science Letters, 455, 85

\bibitem[{Brasser {et~al.}(2020)Brasser, Werner, \& Mojzsis}]{brasser2020impact}
Brasser, R., Werner, S., \& Mojzsis, S. 2020, Icarus, 338, 113514

\bibitem[{Canup(2012)}]{canup2012forming}
Canup, R.~M. 2012, Science, 338, 1052

\bibitem[{Carter {et~al.}(2020)Carter, Lock, \& Stewart}]{carter2020energy}
Carter, P.~J., Lock, S.~J., \& Stewart, S.~T. 2020, Journal of Geophysical Research: Planets, 125, e2019JE006042

\bibitem[{Citron \& Stewart(2022)}]{citron2022large}
Citron, R.~I., \& Stewart, S.~T. 2022, arXiv preprint arXiv:2201.09349

\bibitem[{Collins(2016)}]{collins2016isale}
Collins, G. 2016, Figshare, 10, m9

\bibitem[{Collins(2014)}]{collins_numerical_2014}
Collins, G.~S. 2014, Journal of Geophysical Research: Planets, 119, 2600, \dodoi{10.1002/2014JE004708}

\bibitem[{Collins \& Melosh(2014)}]{collins2014improvements}
Collins, G.~S., \& Melosh, H.~J. 2014, in 45th lunar and planetary science conference, Vol. 2664

\bibitem[{Collins {et~al.}(2004)Collins, Melosh, \& Ivanov}]{collins_modeling_2004}
Collins, G.~S., Melosh, H.~J., \& Ivanov, B.~A. 2004, Meteoritics \& Planetary Science, 39, 217, \dodoi{10.1111/j.1945-5100.2004.tb00337.x}

\bibitem[{Collins {et~al.}(2011)Collins, Melosh, \& Wünnemann}]{collins_improvements_2011}
Collins, G.~S., Melosh, H.~J., \& Wünnemann, K. 2011, International Journal of Impact Engineering, 38, 434, \dodoi{10.1016/j.ijimpeng.2010.10.013}

\bibitem[{{\'C}uk \& Stewart(2012)}]{cuk2012making}
{\'C}uk, M., \& Stewart, S.~T. 2012, science, 338, 1047

\bibitem[{Dahl \& Stevenson(2010)}]{dahl2010turbulent}
Dahl, T.~W., \& Stevenson, D.~J. 2010, Earth and Planetary Science Letters, 295, 177

\bibitem[{Davies(2009)}]{davies2009effect}
Davies, G.~F. 2009, Earth and Planetary Science Letters, 287, 513

\bibitem[{Davison {et~al.}(2022{\natexlab{a}})Davison, Baijal, \& Collins}]{davison2022highMet}
Davison, T., Baijal, N., \& Collins, G. 2022{\natexlab{a}}, in 85th Annual Meeting of The Meteoritical Society, 6362

\bibitem[{Davison {et~al.}(2022{\natexlab{b}})Davison, Baijal, \& Collins}]{davison2022highLPSC}
Davison, T., Baijal, N., \& Collins, G. 2022{\natexlab{b}}, in 53rd Lunar and Planetary Science Conference, Vol. 2695, 2444

\bibitem[{Davison {et~al.}(2016)Davison, Collins, \& Bland}]{davison2016mesoscale}
Davison, T.~M., Collins, G.~S., \& Bland, P.~A. 2016, The Astrophysical Journal, 821, 68

\bibitem[{Day {et~al.}(2016)Day, Brandon, \& Walker}]{day2016highly}
Day, J.~M., Brandon, A.~D., \& Walker, R.~J. 2016, Reviews in Mineralogy and Geochemistry, 81, 161

\bibitem[{Deguen {et~al.}(2014)Deguen, Landeau, \& Olson}]{deguen2014turbulent}
Deguen, R., Landeau, M., \& Olson, P. 2014, Earth and Planetary Science Letters, 391, 274

\bibitem[{Elbeshausen {et~al.}(2009)Elbeshausen, W{\"u}nnemann, \& Collins}]{elbeshausen2009scaling}
Elbeshausen, D., W{\"u}nnemann, K., \& Collins, G.~S. 2009, Icarus, 204, 716

\bibitem[{Elkins-Tanton(2008)}]{elkins2008linked}
Elkins-Tanton, L.~T. 2008, Earth and Planetary Science Letters, 271, 181

\bibitem[{Elkins-Tanton(2012)}]{elkins2012magma}
---. 2012, Annual Review of Earth and Planetary Sciences, 40, 113

\bibitem[{Emsenhuber {et~al.}(2018)Emsenhuber, Jutzi, \& Benz}]{emsenhuber2018sph}
Emsenhuber, A., Jutzi, M., \& Benz, W. 2018, Icarus, 301, 247

\bibitem[{Ganne \& Feng(2017)}]{ganne2017primary}
Ganne, J., \& Feng, X. 2017, Geochemistry, Geophysics, Geosystems, 18, 872

\bibitem[{Genda {et~al.}(2017)Genda, Brasser, \& Mojzsis}]{genda2017terrestrial}
Genda, H., Brasser, R., \& Mojzsis, S. 2017, Earth and Planetary Science Letters, 480, 25

\bibitem[{Gillmann {et~al.}(2016)Gillmann, Golabek, \& Tackley}]{gillmann2016effect}
Gillmann, C., Golabek, G.~J., \& Tackley, P.~J. 2016, Icarus, 268, 295

\bibitem[{Herzberg {et~al.}(2010)Herzberg, Condie, \& Korenaga}]{herzberg2010thermal}
Herzberg, C., Condie, K., \& Korenaga, J. 2010, Earth and Planetary Science Letters, 292, 79

\bibitem[{Ichikawa {et~al.}(2010)Ichikawa, Labrosse, \& Kurita}]{ichikawa2010direct}
Ichikawa, H., Labrosse, S., \& Kurita, K. 2010, Journal of Geophysical Research: Solid Earth, 115

\bibitem[{Itcovitz {et~al.}(2022)Itcovitz, Rae, Citron, Stewart, Sinclair, Rimmer, \& Shorttle}]{itcovitz2022reduced}
Itcovitz, J., Rae, A., Citron, R., {et~al.} 2022, Planetary Science Journal

\bibitem[{Ivanov {et~al.}(1997)Ivanov, Deniem, \& Neukum}]{ivanov_implementation_1997}
Ivanov, B.~A., Deniem, D., \& Neukum, G. 1997, International Journal of Impact Engineering, 20, 411, \dodoi{10.1016/S0734-743X(97)87511-2}

\bibitem[{Jackson \& Wyatt(2012)}]{jackson2012debris}
Jackson, A.~P., \& Wyatt, M.~C. 2012, Monthly Notices of the Royal Astronomical Society, 425, 657

\bibitem[{Jacobson {et~al.}(2014)Jacobson, Morbidelli, Raymond, O'Brien, Walsh, \& Rubie}]{jacobson2014highly}
Jacobson, S.~A., Morbidelli, A., Raymond, S.~N., {et~al.} 2014, Nature, 508, 84

\bibitem[{Jones {et~al.}(2003)Jones, Neal, \& Ely}]{jones2003signatures}
Jones, J.~H., Neal, C.~R., \& Ely, J.~C. 2003, Chemical Geology, 196, 5

\bibitem[{Katsura {et~al.}(2010)Katsura, Yoneda, Yamazaki, Yoshino, \& Ito}]{katsura2010adiabatic}
Katsura, T., Yoneda, A., Yamazaki, D., Yoshino, T., \& Ito, E. 2010, Physics of the Earth and Planetary Interiors, 183, 212

\bibitem[{Kendall \& Melosh(2016)}]{kendall2016differentiated}
Kendall, J.~D., \& Melosh, H. 2016, Earth and Planetary Science Letters, 448, 24

\bibitem[{Kimura {et~al.}(1992)Kimura, Tsuchiyama, Fukuoka, \& Iimura}]{kimura1992antarctic}
Kimura, M., Tsuchiyama, A., Fukuoka, T., \& Iimura, Y. 1992, Antarctic Meteorite Research, 5, 165

\bibitem[{Korenaga(2008)}]{korenaga2008urey}
Korenaga, J. 2008, Reviews of Geophysics, 46

\bibitem[{Korenaga(2021)}]{korenaga2021hadean}
---. 2021, Precambrian Research, 359, 106178

\bibitem[{Korenaga \& Marchi(2023)}]{korenaga2023vestiges}
Korenaga, J., \& Marchi, S. 2023, Proceedings of the National Academy of Sciences, 120, e2309181120

\bibitem[{Kraus {et~al.}(2015)Kraus, Root, Lemke, Stewart, Jacobsen, \& Mattsson}]{kraus2015impact}
Kraus, R.~G., Root, S., Lemke, R.~W., {et~al.} 2015, Nature Geoscience, 8, 269

\bibitem[{Kuwahara \& Sugita(2015)}]{kuwahara2015molecular}
Kuwahara, H., \& Sugita, S. 2015, Icarus, 257, 290

\bibitem[{Landeau {et~al.}(2014)Landeau, Deguen, \& Olson}]{landeau2014experiments}
Landeau, M., Deguen, R., \& Olson, P. 2014, Journal of Fluid Mechanics, 749, 478

\bibitem[{Landeau {et~al.}(2021)Landeau, Deguen, Phillips, Neufeld, Lherm, \& Dalziel}]{landeau2021metal}
Landeau, M., Deguen, R., Phillips, D., {et~al.} 2021, Earth and Planetary Science Letters, 564, 116888

\bibitem[{Lebrun {et~al.}(2013)Lebrun, Massol, Chassefi{\`e}re, Davaille, Marcq, Sarda, Leblanc, \& Brandeis}]{lebrun2013thermal}
Lebrun, T., Massol, H., Chassefi{\`e}re, E., {et~al.} 2013, Journal of Geophysical Research: Planets, 118, 1155

\bibitem[{Leinhardt \& Stewart(2011)}]{leinhardt2011collisions}
Leinhardt, Z.~M., \& Stewart, S.~T. 2011, The Astrophysical Journal, 745, 79

\bibitem[{Lichtenberg {et~al.}(2021)Lichtenberg, Bower, Hammond, Boukrouche, Sanan, Tsai, \& Pierrehumbert}]{lichtenberg2021vertically}
Lichtenberg, T., Bower, D.~J., Hammond, M., {et~al.} 2021, Journal of Geophysical Research: Planets, e2020JE006711

\bibitem[{Lock \& Stewart(2017)}]{lock2017structure}
Lock, S.~J., \& Stewart, S.~T. 2017, Journal of Geophysical Research: Planets, 122, 950

\bibitem[{Lock {et~al.}(2018)Lock, Stewart, Petaev, Leinhardt, Mace, Jacobsen, \& Cuk}]{lock2018origin}
Lock, S.~J., Stewart, S.~T., Petaev, M.~I., {et~al.} 2018, Journal of Geophysical Research: Planets, 123, 910

\bibitem[{Maas {et~al.}(2021)Maas, Manske, W{\"u}nnemann, \& Hansen}]{maas2021fate}
Maas, C., Manske, L., W{\"u}nnemann, K., \& Hansen, U. 2021, Earth and Planetary Science Letters, 554, 116680

\bibitem[{Mann {et~al.}(2012)Mann, Frost, Rubie, Becker, \& Aud{\'e}tat}]{mann2012partitioning}
Mann, U., Frost, D.~J., Rubie, D.~C., Becker, H., \& Aud{\'e}tat, A. 2012, Geochimica et Cosmochimica Acta, 84, 593

\bibitem[{Manske {et~al.}(2021)Manske, Marchi, Plesa, \& Wünnemann}]{manske_impact_2021}
Manske, L., Marchi, S., Plesa, A.-C., \& Wünnemann, K. 2021, Icarus, 357, 114128, \dodoi{10.1016/j.icarus.2020.114128}

\bibitem[{Marchi {et~al.}(2016)Marchi, Black, Elkins-Tanton, \& Bottke}]{marchi2016massive}
Marchi, S., Black, B.~A., Elkins-Tanton, L.~T., \& Bottke, W.~F. 2016, Earth and Planetary Science Letters, 449, 96

\bibitem[{Marchi {et~al.}(2014)Marchi, Bottke, Elkins-Tanton, Bierhaus, Wuennemann, Morbidelli, \& Kring}]{marchi2014widespread}
Marchi, S., Bottke, W., Elkins-Tanton, L., {et~al.} 2014, Nature, 511, 578

\bibitem[{Marchi {et~al.}(2018)Marchi, Canup, \& Walker}]{marchi2018heterogeneous}
Marchi, S., Canup, R., \& Walker, R. 2018, Nature geoscience, 11, 77

\bibitem[{Marchi {et~al.}(2023)Marchi, Rufu, \& Korenaga}]{marchi2023long}
Marchi, S., Rufu, R., \& Korenaga, J. 2023, Nature Astronomy, 1

\bibitem[{Marchi {et~al.}(2020)Marchi, Walker, \& Canup}]{marchi2020compositionally}
Marchi, S., Walker, R.~J., \& Canup, R.~M. 2020, Science advances, 6, eaay2338

\bibitem[{Marinova {et~al.}(2011)Marinova, Aharonson, \& Asphaug}]{marinova2011geophysical}
Marinova, M.~M., Aharonson, O., \& Asphaug, E. 2011, Icarus, 211, 960

\bibitem[{Matthews {et~al.}(2016)Matthews, Shorttle, \& Maclennan}]{matthews2016temperature}
Matthews, S., Shorttle, O., \& Maclennan, J. 2016, Geochemistry, Geophysics, Geosystems, 17, 4725

\bibitem[{Melosh {et~al.}(1992)Melosh, Ryan, \& Asphaug}]{melosh_dynamic_1992}
Melosh, H.~J., Ryan, E.~V., \& Asphaug, E. 1992, Journal of Geophysical Research: Planets, 97, 14735, \dodoi{10.1029/92JE01632}

\bibitem[{Nakajima {et~al.}(2021)Nakajima, Golabek, W{\"u}nnemann, Rubie, Burger, Melosh, Jacobson, Manske, \& Hull}]{nakajima2021scaling}
Nakajima, M., Golabek, G.~J., W{\"u}nnemann, K., {et~al.} 2021, Earth and Planetary Science Letters, 568, 116983

\bibitem[{Neumann {et~al.}(2012)Neumann, Breuer, \& Spohn}]{neumann2012differentiation}
Neumann, W., Breuer, D., \& Spohn, T. 2012, Astronomy \& Astrophysics, 543, A141

\bibitem[{Nikolaou {et~al.}(2019)Nikolaou, Katyal, Tosi, Godolt, Grenfell, \& Rauer}]{nikolaou2019factors}
Nikolaou, A., Katyal, N., Tosi, N., {et~al.} 2019, The Astrophysical Journal, 875, 11

\bibitem[{Nimmo {et~al.}(2008)Nimmo, Hart, Korycansky, \& Agnor}]{nimmo2008implications}
Nimmo, F., Hart, S., Korycansky, D., \& Agnor, C. 2008, Nature, 453, 1220

\bibitem[{O'neill {et~al.}(1995)O'neill, Dingwell, Borisov, Spettel, \& Palme}]{oneil1995experimental}
O'neill, H. S.~C., Dingwell, D., Borisov, A., Spettel, B., \& Palme, H. 1995, Chemical Geology, 120, 255

\bibitem[{Pierazzo \& Melosh(2000)}]{pierazzo2000melt}
Pierazzo, E., \& Melosh, H. 2000, Icarus, 145, 252

\bibitem[{Pierazzo {et~al.}(1997)Pierazzo, Vickery, \& Melosh}]{pierazzo1997reevaluation}
Pierazzo, E., Vickery, A., \& Melosh, H. 1997, Icarus, 127, 408

\bibitem[{Qaddah {et~al.}(2019)Qaddah, Monteux, Clesi, Bouhifd, \& Le~Bars}]{qaddah2019dynamics}
Qaddah, B., Monteux, J., Clesi, V., Bouhifd, M.~A., \& Le~Bars, M. 2019, Physics of the Earth and Planetary Interiors, 289, 75

\bibitem[{Raymond {et~al.}(2013)Raymond, Schlichting, Hersant, \& Selsis}]{raymond2013dynamical}
Raymond, S.~N., Schlichting, H.~E., Hersant, F., \& Selsis, F. 2013, Icarus, 226, 671

\bibitem[{Rubie {et~al.}(2003)Rubie, Melosh, Reid, Liebske, \& Righter}]{rubie2003mechanisms}
Rubie, D., Melosh, H., Reid, J., Liebske, C., \& Righter, K. 2003, Earth and Planetary Science Letters, 205, 239

\bibitem[{Rubie {et~al.}(2016)Rubie, Laurenz, Jacobson, Morbidelli, Palme, Vogel, \& Frost}]{rubie2016highly}
Rubie, D.~C., Laurenz, V., Jacobson, S.~A., {et~al.} 2016, Science, 353, 1141

\bibitem[{Rubie {et~al.}(2015)Rubie, Jacobson, Morbidelli, O’Brien, Young, de~Vries, Nimmo, Palme, \& Frost}]{rubie2015accretion}
Rubie, D.~C., Jacobson, S.~A., Morbidelli, A., {et~al.} 2015, Icarus, 248, 89

\bibitem[{Rudge {et~al.}(2010)Rudge, Kleine, \& Bourdon}]{rudge2010broad}
Rudge, J.~F., Kleine, T., \& Bourdon, B. 2010, Nature Geoscience, 3, 439

\bibitem[{Samuel(2012)}]{samuel2012re}
Samuel, H. 2012, Earth and Planetary Science Letters, 313, 105

\bibitem[{Schaefer \& Fegley(2010)}]{schaefer2010chemistry}
Schaefer, L., \& Fegley, B. 2010, Icarus, 208, 438

\bibitem[{Shinjo \& Umemura(2010)}]{shinjo2010simulation}
Shinjo, J., \& Umemura, A. 2010, International journal of multiphase Flow, 36, 513

\bibitem[{Shoemaker(1962)}]{shoemaker1962interpretation}
Shoemaker, E.~M. 1962, Physics and Astronomy of the Moon, 283

\bibitem[{Simon \& Glatzel(1929)}]{simon1929bemerkungen}
Simon, F., \& Glatzel, G. 1929, Zeitschrift f{\"u}r anorganische und allgemeine Chemie, 178, 309

\bibitem[{Sleep {et~al.}(1989)Sleep, Zahnle, Kasting, \& Morowitz}]{sleep1989annihilation}
Sleep, N.~H., Zahnle, K.~J., Kasting, J.~F., \& Morowitz, H.~J. 1989, Nature, 342, 139

\bibitem[{Sleep {et~al.}(2014)Sleep, Zahnle, \& Lupu}]{sleep2014terrestrial}
Sleep, N.~H., Zahnle, K.~J., \& Lupu, R.~E. 2014, Philosophical Transactions of the Royal Society A: Mathematical, Physical and Engineering Sciences, 372, 20130172

\bibitem[{Stewart(2020)}]{stewart2020equation}
Stewart, S.~T. 2020, Equation of State Model Fe85Si15-ANEOS: Development and documentation (Version SLVTv0.2G1), \url{https://github.com/ststewart/aneos-Fe85Si15-2020},  GitHub

\bibitem[{Svetsov(2005)}]{svetsov2005numerical}
Svetsov, V.~V. 2005, Planetary and Space Science, 53, 1205

\bibitem[{Thompson \& Lauson(1974)}]{thompson_improvements_1974}
Thompson, S.~L., \& Lauson, H.~S. 1974, Improvements in the {CHART} {D} radiation-hydrodynamic code {III}: revised analytic equations of state, Tech. Rep. SC-RR-710714, Sandia National Laboratory, Alberquerque, New Mexico.
\newblock \url{http://inis.iaea.org/Search/search.aspx?orig_q=RN:6209386}

\bibitem[{Tonks \& Melosh(1993)}]{tonks1993magma}
Tonks, W.~B., \& Melosh, H.~J. 1993, Journal of Geophysical Research: Planets, 98, 5319

\bibitem[{Ulvrov{\'a} {et~al.}(2011)Ulvrov{\'a}, Coltice, Ricard, Labrosse, Dubuffet, Vel{\'\i}msk{\`y}, \& {\v{S}}r{\'a}mek}]{ulvrova2011compositional}
Ulvrov{\'a}, M., Coltice, N., Ricard, Y., {et~al.} 2011, Geochemistry, Geophysics, Geosystems, 12

\bibitem[{Villermaux(2007)}]{villermaux2007fragmentation}
Villermaux, E. 2007, Annu. Rev. Fluid Mech., 39, 419

\bibitem[{Villermaux {et~al.}(2004)Villermaux, Marmottant, \& Duplat}]{villermaux2004ligament}
Villermaux, E., Marmottant, P., \& Duplat, J. 2004, Physical review letters, 92, 074501

\bibitem[{Walker(2009)}]{walker2009highly}
Walker, R.~J. 2009, Geochemistry, 69, 101

\bibitem[{Wogan {et~al.}(2023)Wogan, Catling, Zahnle, \& Lupu}]{wogan2023origin}
Wogan, N.~F., Catling, D.~C., Zahnle, K.~J., \& Lupu, R. 2023, The Planetary Science Journal, 4, 169

\bibitem[{Wong(2011)}]{wong2011color}
Wong, B. 2011, nature methods, 8, 441

\bibitem[{Wünnemann {et~al.}(2006)Wünnemann, Collins, \& Melosh}]{wunnemann_strain-based_2006}
Wünnemann, K., Collins, G.~S., \& Melosh, H.~J. 2006, Icarus, 180, 514, \dodoi{10.1016/j.icarus.2005.10.013}

\bibitem[{Wünnemann {et~al.}(2008)Wünnemann, Collins, \& Osinski}]{wunnemann_numerical_2008}
Wünnemann, K., Collins, G.~S., \& Osinski, G.~R. 2008, Earth and Planetary Science Letters, 269, 530, \dodoi{10.1016/j.epsl.2008.03.007}

\bibitem[{Zahnle {et~al.}(1988)Zahnle, Kasting, \& Pollack}]{zahnle1988evolution}
Zahnle, K.~J., Kasting, J.~F., \& Pollack, J.~B. 1988, Icarus, 74, 62

\bibitem[{Zahnle {et~al.}(2020)Zahnle, Lupu, Catling, \& Wogan}]{zahnle2020creation}
Zahnle, K.~J., Lupu, R., Catling, D.~C., \& Wogan, N. 2020, The Planetary Science Journal, 1, 11

\bibitem[{Zimmerman {et~al.}(1999)Zimmerman, Zhang, Kohlstedt, \& Karato}]{zimmerman1999melt}
Zimmerman, M.~E., Zhang, S., Kohlstedt, D.~L., \& Karato, S.-i. 1999, Geophysical Research Letters, 26, 1505

\end{thebibliography}

\begin{acknowledgments}
With thanks to Simone Marchi and the anonymous reviewer for helpful comments. The authors acknowledge \citet{wong2011color} for enabling the creation of accessible graphics. This work was supported by the UK Science and Technology Facilities Council (STFC) grant number ST/T505985/1.. A.S.P.R. gratefully acknowledges funding from Trinity College Cambridge. TMD and GSC were supported by STFC Grant ST/S000615/1. Part of this work used the DiRAC Data Intensive service at Leicester, operated by University of Leicester IT Services, part of the STFC DiRAC HPC Facility (\url{www.dirac.ac.uk}). The equipment was funded by BEIS capital funding via STFC capital Grants ST/K000373/1 and ST/R002363/1 and STFC DiRAC Operations Grant ST/R001014/1. DiRAC is part of the National e-Infrastructure.
\end{acknowledgments}

\appendix
\section{iSALE Modeling} \label{sec:Setup Params} 
iSALE contains a vast number of model parameters and choices available to the user. Tables \ref{tab:Parameters Extended}, \ref{tab:Material} \& \ref{tab:iSALE Asteroid} show the physical, material, and calculation parameters selected for the simulation suite presented in this work. Further detail on each parameter can be found in the iSALE and M-ANEOS manuals (\citet{collins2016isale} and \url{https://github.com/isale-code/M-ANEOS} respectively). In this Section, we highlight some key parameters within our simulation, demonstrating why they are important to modeling the fate of impactor iron or how our choices of parameters affects our results.

\subsection{Simulation Duration} \label{sec:Setup Simulation Duration} 
We run impacts for $3\,\mathrm{hours}$ of model time in our main simulation suite. We find that this duration offers a balance between fully characterizing the accretion behavior of the impactor core material and computational cost. A single $3\,\mathrm{hours}$ simulation running on 30 parallel threads takes approximately half a day to complete, and would take substantially longer if self-consistent gravity were to be included (see Section \ref{sec:Setup Gravity Algorithms}). With the large number of simulations ran for this study across our ranges of impact parameters, minimizing computational cost is desirable. There is also an additional element of storage, with our full data set already occupying $\sim \mathrm{TB}$. 

\begin{figure}[p]
    \centering
    \includegraphics[width=1.0\textwidth]{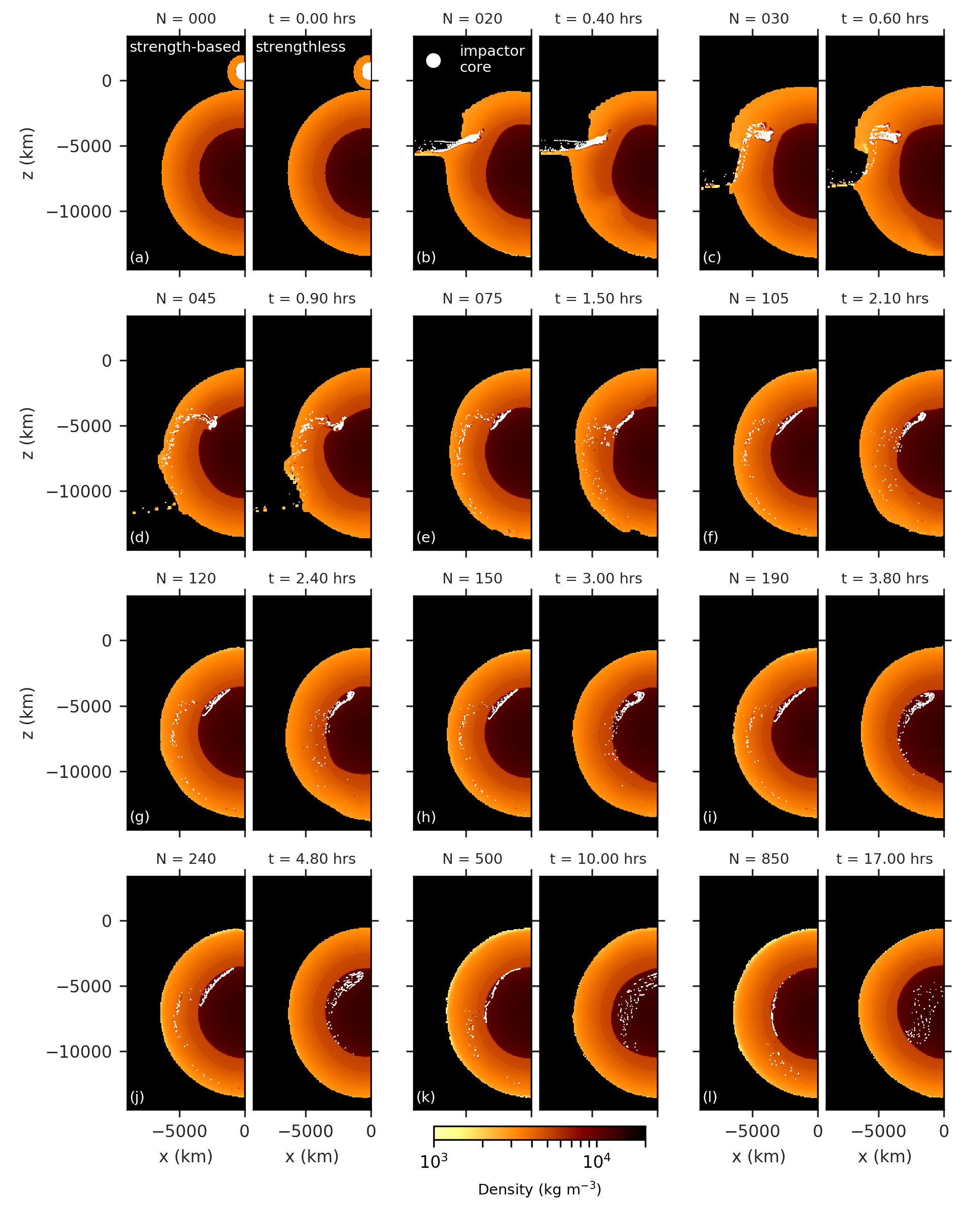}
    \caption{Evolution of models run for $24\,\mathrm{hours}$ with and without material strength in iSALE3D with impact parameters $M_i = 1.93\times10^{22}\,\mathrm{kg}$, $v_i \sim 1.5\,v_{esc}$, $\theta_i = 45^\circ$. Frames are selected to show key stages of the accretion pathways of impactor core material (see text). The Eulerian grid property of cell-averaged material density is shown for the $y=0\,\mathrm{km}$ slice, with tracer particles of the impactor core at all depths (i.e., $y > 0\,\mathrm{km}$ also) overlaid on top (white dots).}
    \label{fig:24 Hour Comparison}
\end{figure}

At the $3\,\mathrm{hour}$ mark, motion of the impactor core material that pertains to our results has approximately ceased in all simulations, meaning that impactor core material that accretes to the planet has reached either the core-mantle boundary (CMB) or the top of the solid mantle. This is valid across the full range of our impact parameters, but is especially so for smaller impactor masses and greater impact angles, where motions of the impactor core material cease well before $3\,\mathrm{hours}$. We now present here the results of a $24\,\mathrm{hours}$ simulation to demonstrate this and justify our choice of simulation time (Figure \ref{fig:24 Hour Comparison}).

As described in Section \ref{sec:Distribution}, there are several key impact stages relevant to the accretion of impactor core material, each of which is highlighted in the panels of Figure \ref{fig:24 Hour Comparison}, including (a) the initial structures of the impactor and planet, (b) the pancaking of the impactor core, (c-e) the accretion of decelerated material at the rear of the impactor core to the planet core as a large clump of molten metal, and (f-h) the dispersal of the impactor core material downstream of the impact site, with the solid mantle acting as a physical barrier to the metal. Note that any motion of impactor core material accreted to the planet core is lateral motion over the surface of the planet core only and does not indicate sinking into the planet core. Such motion may appear as sinking in Figure \ref{fig:24 Hour Comparison} because we plot tracers from all planes (i.e., all planes in the $y$-direction) to show the full accretion of the impactor core, while only plotting the $y=0\,\mathrm{km}$ slice of the background density profile. 

During times spanning $14\,\mathrm{hours}$ after the end of our standard simulation time (Figure \ref{fig:24 Hour Comparison} h-l), we observe no change in the accretion reservoir of impactor core material, with iron coming to rest on top of solid mantle remaining at approximately the same mantle depth (we do not show the full $24\,\mathrm{hours}$ in Figure \ref{fig:24 Hour Comparison}, as rotation of the planet induced by the oblique impact rotates much of the impactor core material out of frame - i.e., $x > 0\,\mathrm{km}$ - an effect that is already visible for some tracers between the final two panels). For impactor core material that accretes to the planet core, we observe its lateral and azimuthal spreading over the core's surface. Regarding the planet mantle itself, it regains its spherical shape well before $t = 3\,\mathrm{hours}$. After this time, we observe that temperature, density, and pressure profiles remain approximately constant, with regions of the mantle that are molten or solid remaining approximately fixed as a result. No material is stationary in its entirety, however, with some residual motion of the planet expressed as small oscillations in the Eulerian grid values. Analysis of the $3/24\,\mathrm{hour}$ simulations, respectively, find accretion mass fractions of impactor core material $f_{core} = 0.379/0.413$, $f_{esc} = 0.041/0.042$, and $f_{mant} = 0.580/0.545$ . 


\subsection{Material Strength} \label{sec:Setup Material Strength} 
Material strength (\texttt{STRESS: 1} in Table \ref{tab:iSALE Asteroid}) is perhaps the most important parameter within the iSALE simulation suite that differentiates our results from previous studies. It is the presence of material strength in the planet mantle that prevents impactor core material from sinking through the mantle directly to the core at azimuthal angles where melting of the mantle down to the core-mantle boundary (CMB) is not achieved. Here, we present an iSALE3D simulation without material strength (i.e., running in ``hydro'' mode) to demonstrate what happens in the absence of strength and to enable direct comparisons. 

Our strengthless simulation is identical to that described in Section \ref{sec:Setup Simulation Duration} (i.e., $24\,\mathrm{hour}$ duration, identical impact parameters) but with \texttt{STRESS} set to zero. The entirety of the planet mantle and planet core are thus fluid at the start of the simulation. The impact follows many of the key impact stages described in Section \ref{sec:Setup Simulation Duration}, with the pancaking of the impactor core and the deceleration of material near its rear leading to a large clump of metal that descends rapidly to the planet core (Figure \ref{fig:24 Hour Comparison}a-d). However, because the mantle is purely a strengthless fluid, the impactor core material that is dispersed downstream of the impact site behaves starkly differently than in the strength-based simulation. Firstly, the metal is more sparsely dispersed at early times, as no solid mantle is present to act as a momentum barrier, meaning that impactor core material continues further along its initial trajectories than the in the strength-based simulations (e-f). Subsequently, because the metal is denser than the surrounding silicates, it simply sinks through the fully molten mantle to the planet core in around $4\,\mathrm{hours}$ (f-i). After this time, impactor core material remains on the surface of the planet core, spreading out laterally and azimuthally with time. We thus find accretion mass fractions of $f_{esc} = 0.051$, $f_{core} = 0.587$, and $f_{mant} = 0.362$, with $f_{mant}$ stemming entirely from vaporized material.

Lastly, without solid material to dampen them (nor fluid viscosity), oscillations in the planetary density-temperature-pressure profiles stemming from the impact continue throughout our simulation time, hence the non-spherical shape of the planet in the strengthless simulation (Figure \ref{fig:24 Hour Comparison}). However, we find that even such oscillations are not sufficient to suspend impactor core material in the molten mantle. 

Our $24\,\mathrm{hour}$ simulations with and without material strength thus provide a direct comparison of impactor core accretion behavior and conclusively show that the inclusion of material is vital in determining the distribution of material between our defined accretion reservoirs of planet core and planet mantle. While we have only demonstrated this for a single impact scenario, the impact stages we have described, and the differences in behavior between the strength-based and strengthless models apply across our range of impact parameters, and indeed become increasingly relevant in impact scenarios where greater mass fractions of the impactor core accrete to the planet interior (e.g., smaller mass, impact angles closer to head-on).

\subsection{Planet Initial Temperature Profile} \label{sec:Setup Planet Init Temp} 
A related parameter to the importance of material strength is the initial temperature profile of the planet. A greater initial mantle temperature will result in more extensive melting as a result of impact, and thus potentially a greater propensity for impactor core material accretion to the planet core. In our iSALE simulations, the planet's geotherm is calculated by defining a surface temperature (\texttt{T\_SURF} in Table \ref{tab:iSALE Asteroid}) from which a conductive profile is then constructed through the lithosphere (depth \texttt{D\_LITH} and thermal gradient \texttt{DTDZSURF}). Moving deeper in to the mantle, a convective profile is calculated (Figure \ref{fig:Pre-Impact}), leading to a much sharper thermal gradient than with the conductive profile of the lithosphere. The convective profile is limited to a maximum temperature at the material solidus, which iSALE calculates using the Simon approximation \citep{simon1929bemerkungen}. Such a thermal profile is indicated by the object temperature profile parameter \texttt{OBJPROF: CONDCONVCAP}. The same profile parameter is used for the planet core, although now the conductive gradient is calculated across the transition zone between the two materials. 

An important consideration in the design of iSALE's thermal profiles is that they are primarily intended as a numerical operation that ensures the correct pre-shock rheology of materials while maintaining somewhat realistic temperatures; they are not designed to be exact in their replication of known geotherms. For example, the solidus limitation on the convective profile results in a strengthless fully molten material without producing the unrealistic temperature excursions that are seen without such a limitation (particularly in the planet core, where $T > 2\times 10^4\,\mathrm{K}$ can be reached, producing a density inversion and destabilizing the planet). This is desirable as the material rheology is more important than the temperature during propagation of the impact shock, and the fact that the temperature may change substantially through the shock itself. 

However, for planetary scale impacts that result in partial melting of the planet mantle, regions of the mantle that are not heavily shocked during impact may affect results, as their yield strengths and melt fractions can be deterministic in the flow of material through the planet. The initial thermal profile of the mantle is thus potentially important, as warmer profiles (but still sub-solidus) are likely to produce greater melt fractions and enable greater movement of material through the mantle. Here, we present results of such a ``warm'' profile simulation, keeping all other parameters identical, as in Section \ref{sec:Setup Simulation Duration}.

In our main simulation suite, our surface temperature of $293\,\mathrm{K}$, lithospheric thickness of $1\times 10^5\,\mathrm{m}$, and thermal gradient of $1\times 10^{-3}\,\mathrm{K m^{-1}}$ lead to a base-of-lithosphere temperature of $T_l = 1293\,\mathrm{K}$. This depicts a relatively cold mantle, even by present-day observations (e.g., $T_l = 1318^{+44}_{-32}\,\mathrm{^\circ C}$ determined by mid-ocean ridge basalts; \citet{matthews2016temperature}). Earth's mantle during the Hadean was likely hotter than present-day due to remnant heat from formation, with evidence for this in the geological record and inferred models (e.g., \citealt{korenaga2008urey, davies2009effect, herzberg2010thermal, ganne2017primary, korenaga2021hadean}). We thus construct a ``warm'' profile for the pre-impact planet mantle by increasing the lithospheric thermal gradient to $1.5\times 10^{-3}\,\mathrm{K m^{-1}}$, while maintaining a surface temperature of $293\,\mathrm{K}$ to reflect the postulated temperate surface conditions of Earth during the Hadean (Section \ref{sec:Introduction}). We thus obtain a base of lithosphere temperature of $1793\,\mathrm{K}$, which is notably hot compared to estimates and should provide an end-member case to study any changes in behavior during our simulations that come about from such a change in mantle conditions. 

Our test scenario uses impact parameters $M_i = 1.93\times10^{22}\,\mathrm{kg}$, $v_i \sim 1.5\,v_{esc}$, $\theta_i = 45^\circ$, as in Section \ref{sec:Setup Simulation Duration}, representing typical parameters for a large impactor (Section \ref{sec:iSALE3D}). Such an impact scenario is suitable for discussion of this parameter as it exhibits a fine balance between impactor core material accreting to the planet core and the planet mantle, and therefore out of all of our simulations should be among the most strongly affected by changes in the mantle melt fractions induced by changing $T_l$. We find that impactor core material generally settles at a greater depth within the planet mantle with the greater $T_l$. However, there is no substantial changes in the mass fractions of impactor core material that accretes to each planetary reservoir: $f_{core} = 0.379/0.412$, $f_{esc} = 0.041/0.042$, and $f_{mant} = 0.580/0.545$ for the $1293/1793\,\mathrm{K}$ cases respectively. We thus find a $\sim 3\%$ change in favor of accretion to the planet core when the mantle temperature increases (i.e., the expected direction of change). 

Given that our presented model is near the upper end of Hadean mantle temperatures, and that our chosen impact parameters should maximize the difference we observe, we suggest that the effect of base of lithosphere temperature within our iSALE simulations is not substantial across our main suite of simulations. If the mantle geotherm is moved even closer to the material solidus, then we would expect to see greater change. The most extreme case of this would be fixing the geotherm to the solidus, which would ensure a liquid post-impact mantle and remove the effects of material strength within the model (e.g., Section \ref{sec:Setup Material Strength}). Such a scenario thus requires increasing lower mantle temperatures beyond the mantle adiabat (i.e., superadiabatic), which is unlikely over sustained depth within the mantle (e.g., \citealt{katsura2010adiabatic}). The results of our main simulation suite are, therefore, relatively robust over a range of mantle temperatures as long as the planet geotherm producess a fully solid mantle.

\subsection{Gravity Algorithms} \label{sec:Setup Gravity Algorithms} 
iSALE3D being limited to a central gravity scheme, instead a self-consistent time-dependent gravity field, will lead to differences in the gravity field experienced within our Eulerian grid. This difference in acceleration may then potentially lead to changes in the fate of impactor core material. In order to explore these potential differences, we ran additional simulations using iSALE2D, where self-consistent gravity is an available feature (\texttt{GRAD\_TYPE: SELF} in Table \ref{tab:iSALE Asteroid}) and compare results to an equivalent head-on impact in iSALE3D (all iSALE2D simulations are limited to head-on collisions only by nature of the geometry). All parameters in these three simulations are identical except for the gravity algorithm and parameters associated with dimensionality (e.g., boundary conditions).

\begin{figure}[t!]
    \centering
    \includegraphics[width=\textwidth]{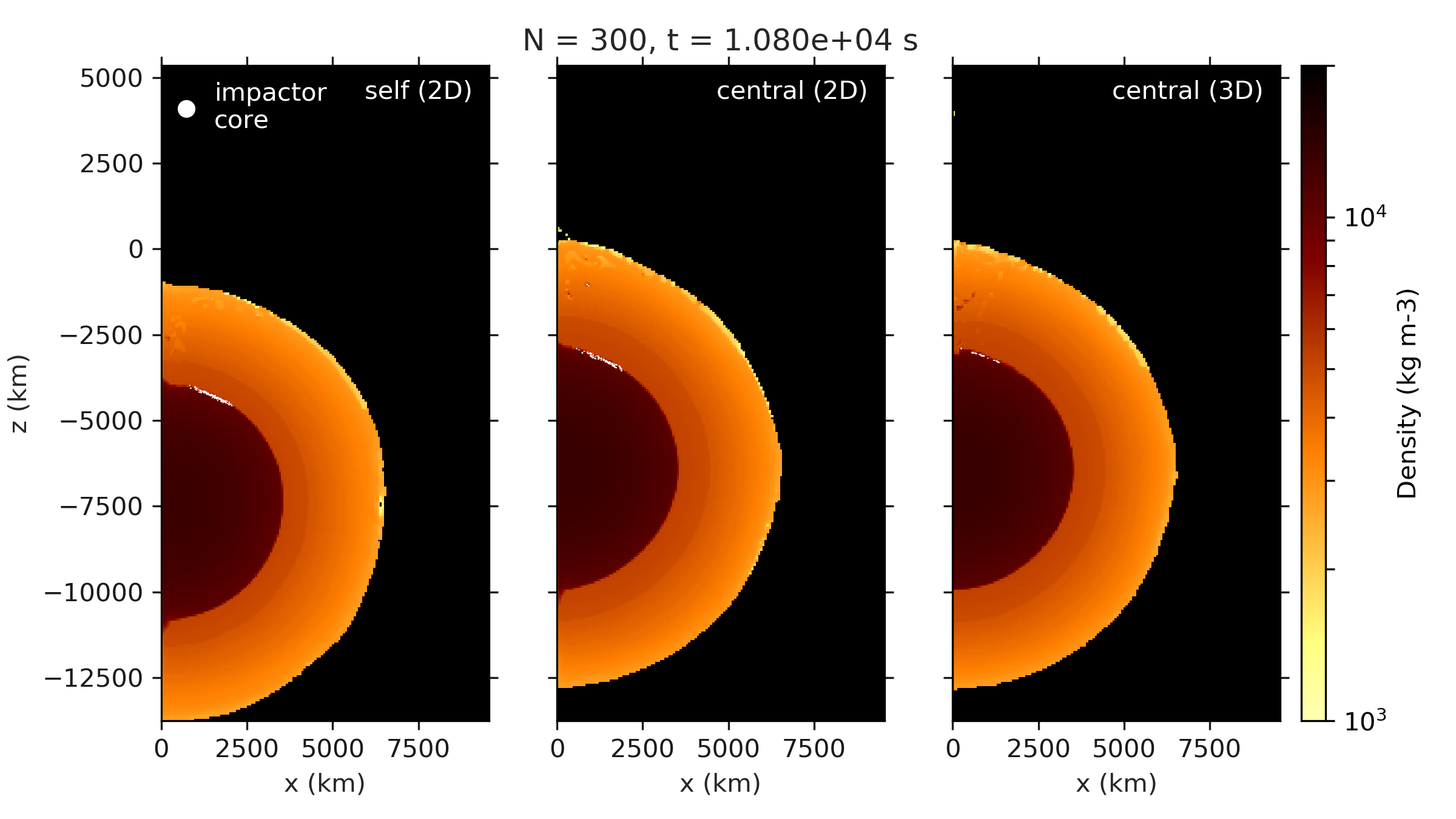}
    \caption{Density profiles at the final timestep of the simulations for iSALE2D simulations under central (\textit{left}) and self (\textit{center}) gravity, as well an iSALE3D simulation under central gravity (\textit{right}). Lagrangian tracer particles representing impactor core material are shown as white dots, visible almost exclusively at the core-mantle boundary (see text for more details). Impact parameters are $M_i = 1.93\times10^{22}\,\mathrm{kg}$, $v_i \sim 1.5\,v_{esc}$, $\theta_i = 0^\circ$. Because of the symmetry of 2D simulations, iSALE models only one hemisphere of the planet slice along $y=0\,\mathrm{km}$ to reduce computation time. iSALE3D models one hemisphere of the full planet for similar reasons of symmetry, but because of the head-on nature of the impact in this specific scenario, the impact is additionally symmetrical in the x-axis, hence we show only one half of the simulated space (\textit{right}).}
    \label{fig:2D 3D Comparison}
\end{figure}

Firstly, we compare the two 2D simulations under self gravity and central gravity (Figure \ref{fig:2D 3D Comparison}a \& b). We find that the distributions of impactor core material are almost identical, with impactor core material congregated in a sheet at the CMB spanning azimuthal angles $15^\circ \sim 35^\circ$ from the central collision axis. The only noticeable difference between the two simulations in terms of impactor core placement is in a single fragment that is transported through the planet mantle by an upwelling piece of the planet core that breaks off during the planet's rebound from the initial impact. Under self gravity, a single Langrangian tracer particle's worth of mass is transported away by this effect, ending up just below the planet surface at $x \sim 3000\,\mathrm{km}$ (Figure \ref{fig:2D 3D Comparison}). Under central gravity, two tracer particles' worth of material is transported away and ends up more centrally located in the planet mantle, although we predict that the upwelling material would eventually move azimuthally once near the surface. Such differences speak to the small-scale differences in the structures of plumes that we generally observe between the two gravity algorithms. Out of the 2640 tracer particles of the impactor core for this size of impactor, a single particle's difference is an insubstantial change between the two algorithms.

The most noticeable difference between the two gravity regimes is the vertical shift between the two bodies' final positions, despite all bodies starting in the same locations between all simulations. This is due to the planet being able to move away from its initial center of mass under self gravity, converting the momentum of the impactor into momentum of the planet. Such a shift in the planet's center of mass is not possible under central gravity by definition. Instead, the kinetic energy is converted into producing a stronger rebound of the planet mantle in the aftermath of crater formation, resulting in a more energetic and thus taller central peak along the impact axis. This primarily affects material that goes on to be resident near the surface of the planet when the ejecta peak collapses, however, and the magnitude of the effect diminishes at locations closer to the planet core. 

The differences in central peak heights thus do not result in differences in the accretion of impactor core material, as the molten metal penetrates deeply into the planet, reaching the lower mantle during crater formation. Indeed, when the central upwelling occurs during mantle rebound, impactor core material is only lifted marginally higher above the CMB under central gravity than self gravity. The material then rapidly descends again to the CMB in both cases once the rebounding motions of the mantle have diminished. Such behavior falls into the broader trend of small-scale differences that we observe in temperature-density-pressure fields between the two gravity regimes, as already highlighted with regards to tracer particles that we find in the mantle at the end of our simulations. These differences lead to perturbations in individual structures (e.g., rising plumes of high-temperature lower mantle material, Figure \ref{fig:2D 3D Comparison}) but do not affect the larger scale dynamic behavior of the molten mantle. We thus find that, in our 2D impact scenario, the different gravity regimes have no effect on the overall accretion patterns of the impactor core material (i.e., where it accretes to the impactor core or the solid mantle), but rather affect how each individual fluid parcel ends up in the final location that it does.  

Comparing the 2D simulation under self gravity with the 3D simulation (central gravity), we find that the distribution of impactor core material is again almost identical (Figure \ref{fig:2D 3D Comparison}), with the vast majority of such material sitting on top of the planet core in a sheet at the end of both simulations. In the 3D simulation, impactor core material can spread laterally upon reaching the surface of the planet core, leading to a thinner sheet than in the 2D simulation and hence fewer tracer particles sitting in the $y\sim 0\,\mathrm{km}$ plane (Figure \ref{fig:2D 3D Comparison}). Lateral spreading may affect our results through the criterion we use to decide when an impactor core tracer particle has reached the planet core (i.e., when the tracer reaches a grid cell with non-zero mass fractions of planet core material). However, the lateral spreading acts to thin out the sheet and thus increase the likelihood of the criterion being met, ensuring that all tracers reaching the planet core are indeed in contact with cells containing material of the planet core. 

The vertical shift between the final states under self gravity and central gravity is still present in the 3D simulation, and by approximately the same amount as in the 2D simulation. (Figure \ref{fig:2D 3D Comparison}). We still observe small-scale changes in the dynamical structures formed within the mantle (i.e., rising plumes of hot material), especially near the central collision axis where the impact energy is focused due to the head-on nature of the collision, but again they do not affect the determined distribution of impactor core material, rather the specific location of the material in its accretion reservoir. The example of the tracer particles that are transported by the broken off pieces of planet core can once more be used to demonstrate this. In the 3D simulation, we find one such tracer particle that originates from the $y=0\,\mathrm{km}$ plane of the impactor. This tracer is not visible in Figure \ref{fig:2D 3D Comparison} as it moves to $y > 0\,\mathrm{km}$ during the simulation, but its path through the mantle is remarkably similar to the similar tracer in the 2D simulation under self gravity, albeit at a marginally lower azimuthal angle from the collision axis. 

We thus find that there is little difference between the three impact regimes (i.e., 2D self gravity, 2D central gravity, and  3D central gravity) in our test scenario. This scenario, through its use of our largest impactor and its production of the largest distortion to the planet's mass distribution, should provide some of the greatest differences in results if they are substantial. Consequently, we suggest that iSALE3D's inability to calculate the planet's gravity field in a time-dependent and self-consistent manner does not substantially affect the results of our study with regards to material accreting to the planet interior.

We would like to highlight, however, that without the full development of self-consistent gravity in iSALE3D, we cannot know the consequences with absolute certainty, as our investigation has not been able to account for impact angles away from head-on collisions. A key difference is that the nature of the $\theta_i = 0^\circ$ collisions results in impactor core material almost entirely accreting to the planet core (i.e., $f_{core} \sim 1.0$) in all cases. Thus, while the small changes to the thermal structures we observe in our comparisons (Figure \ref{fig:2D 3D Comparison}) have little consequence to the large-scale accretion patterns of the impactor core material, such small changes may have greater consequence in more oblique impacts, where small changes to a plume's dynamics may carry a small piece of the impactor core from a path accreting to the planet core to one that ends up at a region of solid mantle. The head-on nature of 2D simulations additionally means that we are unable to fully explore the consequences of different gravity algorithms on impactor core material that escapes the planet. For the planet interior, however, we suggest that the larger scale accretion patterns should not be substantially affected, and we propose that such effects will likely result in percent-level changes in the mass fractions of impactor core material accreting to each reservoir.

\begin{table}[h!]
    \centering
    \caption{iSALE simulation parameters used across the extended suite of simulations, as described in Section \ref{sec:param_applicable}.}
    \label{tab:Parameters Extended}
    \begin{tabular}{rlllllll}
        \toprule
        Parameter  &  Units   &  Target  &  Small  &  Regular  &  Medium  &  Large  & Huge  \\
        \midrule
        Mass  &  kg  &  $5.755\times10^{24}$  &  $2.775\times10^{21}$  &  $5.103\times10^{21}$  &  $8.379\times10^{21}$  &  $3.604\times10^{22}$  &  $1.748\times10^{23}$ \\
        \quad  &  $M_\Earth$  &  1.0  &  0.0005  &  0.0009  &  0.0015  & 0.0063  &  0.0304  \\
        \midrule
        No. of Cells (Mantle + Core)  &  -     &  200    &  18    &  22     &  26    &  42    &  70    \\
        Diameter (Mantle + Core)      &  km    &  12740  &  1147  &  1401   &  1656  &  2675  &  4459  \\
        \midrule
        No. of Cells (Core)           &  -     &  110    &  10    &  12     &  14    &  22    &  40    \\
        Diameter (Core)               &  km    &  7007   &  637   &  764    &  892   &  1401  &  2548  \\
        \midrule
        Core-Mantle Mass Ratio        & -      &  0.366  &  0.259 &  0.274  &  0.255 &  0.244 & 0.310  \\ 
        \midrule
        Tracers in Impactor Core      & -      &  -      &  216   &  419    &  640   &  2640   &  13940  \\
        \bottomrule
    \end{tabular}

    \bigskip
    \bigskip
    
    \centering
    \caption{Material setup parameters for mantle and core materials.}
    \label{tab:Material}
    \small{
    \begin{verbatim}
                        ------------------------------------------------------------
                        Parameter  Description            Material 1    Material 2
                        ------------------------------------------------------------
                        MATNAME    Material name          : mant__      : core__   
                        EOSNAME    EOS name               : forstrm     : ironS20  
                        EOSTYPE    EOS type               : aneos       : aneos         
                        STRMOD     Strength model         : ROCK        : VNMS          
                        DAMMOD     Damage model           : IVANOV      : NONE          
                        ACFL       Acoustic fluidisation  : NONE        : NONE          
                        PORMOD     Porosity model         : NONE        : NONE          
                        THSOFT     Thermal softening      : OHNAKA      : OHNAKA          
                        LDWEAK     Low density weakening  : POLY        : POLY          
                        ------------------------------------------------------------
                        ANETABLE   make or read           : MAKE        : MAKE          
                        ANEND      no. of density nodes   : 240         : 240       
                        ANENT      no. of thermal nodes   : 120         : 120        
                        --------general parameters ---------------------------------
                        POIS       pois                   : 2.5D-01     : 3.0D-01  
                        --------thermal parameters ---------------------------------
                        TMELT0     tmelt0                 : 1.373D+03   : 1.811D+03
                        TFRAC      tfrac                  : 2.0D+00     : 1.2D+00  
                        ASIMON     a_simon                : 1.52D+09    : 6.D+09   
                        CSIMON     c_simon                : 4.05D+00    : 3.D+00   
                        ---------shear strength of intact material -----------------
                        YINT0      yint0                  : 1.D+07      : 1.D+08   
                        FRICINT    fricint                : 1.2D+00     : XXXXXX
                        YLIMINT    ylimint                : 3.5D+09     : XXXXXX 
                        ---------shear strength of damaged material ----------------
                        YDAM0      ydam0 (ycoh)           : 1.D+04      : 1.D+04       
                        FRICDAM    fricdam                : 6.D-01      : 4.D-01        
                        YLIMDAM    ylimdam                : 3.5D+09     : 2.5D+09  
                        ---------damage parameters----------------------------------
                        IVANOV_A   Damage parameter       : 1.0000D-04  : 1.0000D-04
                        IVANOV_B   Damage parameter       : 1.0000D-11  : 1.0000D-11
                        IVANOV_C   Damage parameter       : 3.0000D+08  : 3.0000D+08
                        ------------------------------------------------------------
                        PMININ     minimum pressure       : XXXXXX      : -1.750D+08
                        ------------------------------------------------------------ 
    \end{verbatim}
    }
\end{table} 

\begin{table}[h!]
    \centering
    \caption{Grid, impactor, and target setup parameters for the iSALE simulations.}
    \label{tab:iSALE Asteroid}
    \small{
    \begin{verbatim}
            ------------------- Mesh Geometry Parameters ---------------------------
            GRIDH       horizontal cells         : 0          : 340         : 0
            GRIDV       vertical cells           : 0          : 280         : 0
            GRIDD       depth cells              : 0          : 140         : 0
            GRIDEXT     ext. factor              : 1.06d0
            GRIDSPC     grid spacing             : 63.7D3
            GRIDSPCM    max. grid spacing        : -20.D0
            SYMMETRY    grid symmetry            : 2
            LP_TOLER    pressure tolerance       : 1.D-6
            ------------------- Global Setup Parameters ----------------------------
            S_TYPE      setup type               : PLANET
            T_SURF      surface temp             : 293.D0
            DTDZSURF    temp. grad. surf.        : 10.D-3
            D_LITH      lithosp. thickness       : 100.D3
            R_PLANET    planet radius            : 6370.D3
            GRAD_TYPE   gradient type            : CENTRAL
            GRAD_DIM    gradient dimension       : 3
            ROCUTOFF    density cutoff           : 5.D0
            ------------------- Projectile Parameters ------------------------------
            OBJNUM      number of objects        : 4
            PROJNUM     number of projectiles    : 2
            PR_TRACE    collision tracers        : 1
            OBJRESH     CPPR horizontal          : 11         : 21          : 55          : 100
            OBJRESV     CPPR vertical            : 11         : 21          : 55          : 100
            OBJRESD     CPPR depth               : 11         : 21          : 55          : 100
            OBJVEL      object velocity          : -16.0D3    : -16.0D3     : 0.D0        : 0.D0
            ANGLE       inc. angle (X-Z)         : 45.D0      : 45.D0       : 0.D0        : 0.D0
            ANGLE2      inc. angle (Y-Z)         : 00.D0      : 00.D0       : 00.D0       : 00.D0
            OBJMAT      object material          : core_i     : mant_i      : core_t      : mant_t
            OBJTYPE     object type              : SPHEROID   : SPHEROID    : SPHEROID    : SPHEROID
            OBJTPROF    object temp prof         : CONST      : CONDCONVCAP : CONDCONVCAP : CONDCONVCAP
            OBJTEMP     object temp              : 2200.D0    : 293.D0      : 2200.D0     : 293.D0
            OBJOFF_H    object shift hor         : 65         : 65          : 65          : 65
            OBJOFF_V    object shift ver         : 227        : 217         : 62          : 17
            OBJOFF_D    object shift dpth        : 0          : 0           : 0           : 0
            ------------------- Time Parameters ------------------------------------
            DT          initial time increment   : 1.0D-3
            DTMAX       maximum timestep         : 10.0D-1
            TEND        end time                 : 10800.D0
            DTSAVE      save interval            : 3600.D-2
            ------------------- Boundary Conditions --------------------------------
            BND_L       left                     : OUTFLOW
            BND_R       right                    : OUTFLOW
            BND_B       bottom                   : OUTFLOW
            BND_T       top                      : OUTFLOW
            BND_F       front                    : FREESLIP
            BND_BK      back                     : OUTFLOW
            ------------------- Tracer Particle Parameters -------------------------
            TR_QUAL     integration qual.        : 1
            TR_SPCH     spacing horiz.           : -1.D0      : -1.D0       : -1.D0
            TR_SPCV     spacing vertical         : -1.D0      : -1.D0       : -1.D0
            TR_SPCD     spacing depth            : -1.D0      : -1.D0       : -1.D0
            TR_VAR      add. tracer fields       : #TrP-Trp-TrT-Trt-TrE-TrM-Tru-Trv-Trw#
            TR_MOTION   moving algorithm         : MATERIAL
            ------------------- Data Saving Parameters -----------------------------
            VARLIST     list of variables        : #Den-Pre-Tmp-VEL-Yld#
            -------------------- Superflous later ----------------------------------
            STRESS      calc_stress              : 1
    \end{verbatim}
    }
\end{table}
\clearpage
\section{Iron Distribution Results} \label{sec:Results Table}
\begin{figure}[h!]
    \centering
    \includegraphics[width=0.75\textwidth]{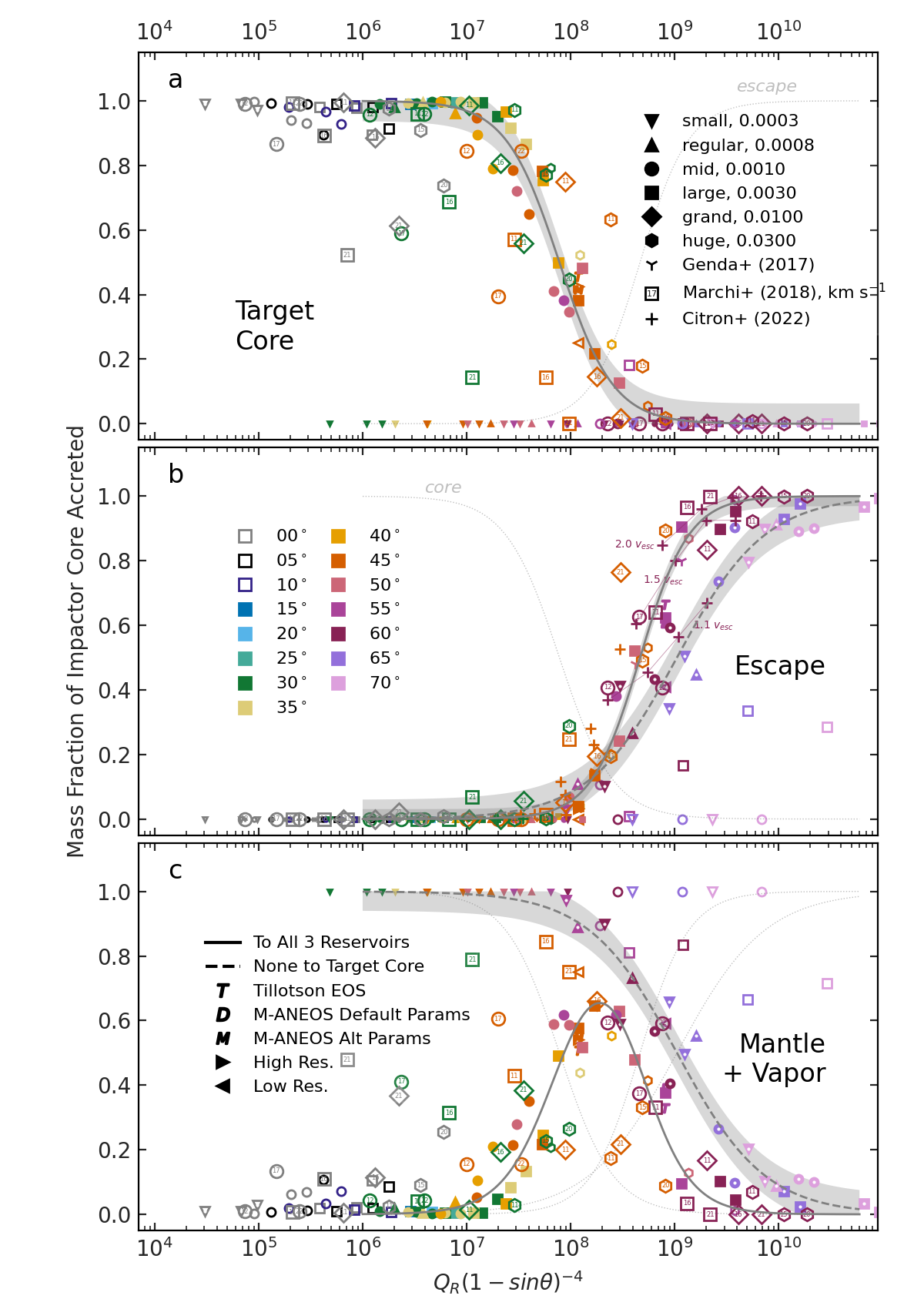}
    \caption{Alternative version of Figure \ref{fig:Parameterization}, with additional detail given for \citet{marchi2018heterogeneous} and \citet{citron2022large} impact parameters to assist in direct comparison between iSALE simulations and their SPH equivalents with similar impact parameters. Impacts with $M_i/M_t \geq 0.0100$ are also shown, which belong to the \textit{giant} impacts accretion mode and show distinctly different impactor core material distribution as a function of $Q_L$.}
    \label{fig:Parameterization_Full_Detail}
\end{figure}

\begin{table}[h!]  
    \centering
    \caption{Distribution of impactor core material in all iSALE simulations (huge- \& large-sized impactors). Italicized entry indicates the low resolution simulation. Bold entry indicates the high resolution simulation.}
    \label{tab:Results_Large_1}
    \begin{tabular}{c|c|c|cccc}
        \toprule
        Impactor Mass & Impact Velocity & Impact Angle  &  Escape  &  Core   &  Mantle  &  Vapor \\
        (kg)          & (km s$^{-1}$)   & (degrees)     &  \quad   &  \quad  &  \quad   &  \quad \\
        \midrule
        \multirow[c]{6}{*}{1.75e23} & \multirow[c]{6}{*}{16.0} & 30 & 0.00000 & 0.79326 & 0.20043 & 0.00631 \\
        \quad & \quad & 35 & 0.03874 & 0.52181 & 0.40352 & 0.03594 \\
        \quad & \quad & 40 & 0.20000 & 0.24720 & 0.49412 & 0.05868 \\
        \quad & \quad & 45 & 0.53063 & 0.05452 & 0.39053 & 0.02432 \\
        \quad & \quad & 50 & 0.86879 & 0.00294 & 0.11994 & 0.00832 \\
        \quad & \quad & 55 & 0.98795 & 0.00000 & 0.00380 & 0.00825 \\
        \midrule
        \multirow[c]{41}{*}{3.60e22} & \multirow[c]{11}{*}{20.0} & 00 & 0.00000 & 0.89394 & 0.10492 & 0.00114 \\
        \quad & \quad & 05 & 0.00000 & 0.91402 & 0.08333 & 0.00265 \\
        \quad & \quad & 30 & 0.00000 & 0.95189 & 0.03939 & 0.00871 \\
        \quad & \quad & 35 & 0.00000 & 0.86667 & 0.11818 & 0.01515 \\
        \quad & \quad & 40 & 0.00871 & 0.49924 & 0.39924 & 0.09280 \\
        \quad & \quad & 45 & 0.13598 & 0.21856 & 0.33977 & 0.30568 \\
        \quad & \quad & 50 & 0.52045 & 0.00000 & 0.18295 & 0.29659 \\
        \quad & \quad & 55 & 0.90530 & 0.00000 & 0.00758 & 0.08712 \\
        \quad & \quad & 60 & 0.95379 & 0.00000 & 0.00038 & 0.04583 \\
        \quad & \quad & 65 & 0.97689 & 0.00000 & 0.00038 & 0.02273 \\
        \quad & \quad & 70 & 0.99394 & 0.00000 & 0.00038 & 0.00568 \\
        \quad & \multirow[c]{18}{*}{16.0} & 00 & 0.00000 & 0.97765 & 0.02197 & 0.00038 \\
        \quad & \quad & 05 & 0.00000 & 0.98182 & 0.01705 & 0.00114 \\
        \quad & \quad & 10 & 0.00000 & 0.99356 & 0.00455 & 0.00189 \\
        \quad & \quad & 15 & 0.00000 & 0.99091 & 0.00682 & 0.00227 \\
        \quad & \quad & 20 & 0.00000 & 0.99242 & 0.00568 & 0.00189 \\
        \quad & \quad & 25 & 0.00000 & 0.99583 & 0.00265 & 0.00152 \\
        \quad & \quad & 30 & 0.00000 & 0.99508 & 0.00379 & 0.00114 \\
        \quad & \quad & 35 & 0.00000 & 0.91705 & 0.07879 & 0.00417 \\
        \quad & \quad & 40 & 0.00000 & 0.75455 & 0.23144 & 0.01402 \\
        \quad & \quad & \textit{45} & \textit{0.00000} & \textit{0.25000} & \textit{0.69524} & \textit{0.05476} \\
        \quad & \quad & 45 & 0.03674 & 0.37879 & 0.35379 & 0.23068 \\
        \quad & \quad & \textbf{45} & \textbf{0.03268} & \textbf{0.42518} & \textbf{0.30731} & \textbf{0.23483} \\
        \quad & \quad & 50 & 0.24242 & 0.12689 & 0.37045 & 0.26023 \\
        \quad & \quad & 55 & 0.62386 & 0.00038 & 0.18712 & 0.18864 \\
        \quad & \quad & 55 & 0.40952 & 0.00000 & 0.31190 & 0.27857 \\
        \quad & \quad & 60 & 0.89735 & 0.00000 & 0.01894 & 0.08371 \\
        \quad & \quad & 65 & 0.92803 & 0.00000 & 0.00038 & 0.07159 \\
        \quad & \quad & 70 & 0.96667 & 0.00000 & 0.00038 & 0.03295 \\
        \quad & \multirow[c]{12}{*}{11.2} & 00 & 0.00000 & 0.98068 & 0.01932 & 0.00000 \\
        \quad & \quad & 05 & 0.00000 & 0.99091 & 0.00795 & 0.00114 \\
        \quad & \quad & 10 & 0.00000 & 0.98561 & 0.01364 & 0.00076 \\
        \quad & \quad & 30 & 0.00000 & 0.99621 & 0.00189 & 0.00189 \\
        \quad & \quad & 35 & 0.00000 & 0.99508 & 0.00076 & 0.00417 \\
        \quad & \quad & 40 & 0.00000 & 0.96705 & 0.03030 & 0.00265 \\
        \quad & \quad & 45 & 0.00000 & 0.78295 & 0.20341 & 0.01364 \\
        \quad & \quad & 50 & 0.00000 & 0.48220 & 0.41061 & 0.10720 \\
        \quad & \quad & 55 & 0.00833 & 0.18144 & 0.69394 & 0.11629 \\
        \quad & \quad & 60 & 0.16553 & 0.00000 & 0.75076 & 0.08371 \\
        \quad & \quad & 65 & 0.33523 & 0.00000 & 0.59091 & 0.07386 \\
        \quad & \quad & 70 & 0.28561 & 0.00000 & 0.63561 & 0.07879 \\
    \bottomrule
    \end{tabular}
\end{table}

\newpage  
\setcounter{table}{3}
\begin{table}[h!]  
    \centering
    \caption{Distribution of impactor core material in all iSALE simulations (continued, mid- \& regular-size impactors).}
    \label{tab:Results_Large_2}
    \begin{tabular}{c|c|c|cccc}
        \toprule
        Impactor Mass & Impact Velocity & Impact Angle  &  Escape  &  Core   &  Mantle  &  Vapor \\
        (kg)          & (km s$^{-1}$)   & (degrees)     &  \quad   &  \quad  &  \quad   &  \quad \\
        \midrule
        \multirow[c]{36}{*}{8.38e21} & \multirow[c]{12}{*}{20.0} & 00 & 0.00000 & 0.92969 & 0.07031 & 0.00000 \\
        \quad & \quad & 05 & 0.00000 & 0.89375 & 0.10469 & 0.00156 \\
        \quad & \quad & 10 & 0.00000 & 0.92812 & 0.07187 & 0.00000 \\
        \quad & \quad & 30 & 0.00000 & 0.99844 & 0.00156 & 0.00000 \\
        \quad & \quad & 35 & 0.00000 & 0.99687 & 0.00156 & 0.00156 \\
        \quad & \quad & 40 & 0.00000 & 0.79063 & 0.19062 & 0.01875 \\
        \quad & \quad & 45 & 0.00000 & 0.64844 & 0.31250 & 0.03906 \\
        \quad & \quad & 50 & 0.06875 & 0.34531 & 0.43281 & 0.15313 \\
        \quad & \quad & 55 & 0.38125 & 0.00000 & 0.25312 & 0.36562 \\
        \quad & \quad & 60 & 0.59375 & 0.00000 & 0.04844 & 0.35781 \\
        \quad & \quad & 65 & 0.90312 & 0.00000 & 0.00156 & 0.09531 \\
        \quad & \quad & 70 & 0.90000 & 0.00000 & 0.00156 & 0.09844 \\
        \quad & \multirow[c]{12}{*}{16.0} & 00 & 0.00000 & 0.93906 & 0.06094 & 0.00000 \\
        \quad & \quad & 05 & 0.00000 & 0.98906 & 0.01094 & 0.00000 \\
        \quad & \quad & 10 & 0.00000 & 0.96562 & 0.03438 & 0.00000 \\
        \quad & \quad & 30 & 0.00000 & 0.99219 & 0.00781 & 0.00000 \\
        \quad & \quad & 35 & 0.00000 & 0.99531 & 0.00313 & 0.00156 \\
        \quad & \quad & 40 & 0.00000 & 0.89531 & 0.09844 & 0.00625 \\
        \quad & \quad & 45 & 0.00000 & 0.78438 & 0.18125 & 0.03438 \\
        \quad & \quad & 50 & 0.00000 & 0.41094 & 0.45781 & 0.13125 \\
        \quad & \quad & 55 & 0.10625 & 0.00000 & 0.55000 & 0.34375 \\
        \quad & \quad & 60 & 0.43281 & 0.00000 & 0.16406 & 0.40313 \\
        \quad & \quad & 65 & 0.73594 & 0.00000 & 0.02813 & 0.23594 \\
        \quad & \quad & 70 & 0.89062 & 0.00000 & 0.00156 & 0.10781 \\
        \quad & \multirow[c]{12}{*}{11.2} & 00 & 0.00000 & 0.99687 & 0.00313 & 0.00000 \\
        \quad & \quad & 05 & 0.00000 & 0.99219 & 0.00781 & 0.00000 \\
        \quad & \quad & 10 & 0.00000 & 0.98125 & 0.01719 & 0.00156 \\
        \quad & \quad & 30 & 0.00000 & 0.99062 & 0.00937 & 0.00000 \\
        \quad & \quad & 35 & 0.00000 & 0.99375 & 0.00625 & 0.00000 \\
        \quad & \quad & 40 & 0.00000 & 0.99687 & 0.00313 & 0.00000 \\
        \quad & \quad & 45 & 0.00000 & 0.94688 & 0.05000 & 0.00313 \\
        \quad & \quad & 50 & 0.00000 & 0.72188 & 0.25781 & 0.02031 \\
        \quad & \quad & 55 & 0.00000 & 0.38281 & 0.42500 & 0.19219 \\
        \quad & \quad & 60 & 0.00000 & 0.00000 & 0.74375 & 0.25625 \\
        \quad & \quad & 65 & 0.00000 & 0.00000 & 0.73438 & 0.26562 \\
        \quad & \quad & 70 & 0.00000 & 0.00000 & 0.86719 & 0.13281 \\
        \midrule
        \multirow[c]{9}{*}{5.10e21} & \multirow[c]{9}{*}{16.0} & 30 & 0.00000 & 0.98091 & 0.01671 & 0.00239 \\
        \quad & \quad & 35 & 0.00000 & 0.99523 & 0.00477 & 0.00000 \\
        \quad & \quad & 40 & 0.00000 & 0.96181 & 0.03103 & 0.00716 \\
        \quad & \quad & 45 & 0.00000 & 0.00000 & 0.99284 & 0.00716 \\
        \quad & \quad & 50 & 0.00000 & 0.00000 & 0.90215 & 0.09785 \\
        \quad & \quad & 55 & 0.10979 & 0.00000 & 0.62768 & 0.26253 \\
        \quad & \quad & 60 & 0.26730 & 0.00000 & 0.34129 & 0.39141 \\
        \quad & \quad & 65 & 0.44630 & 0.00000 & 0.00477 & 0.54893 \\
        \quad & \quad & 70 & 0.91169 & 0.00000 & 0.00239 & 0.08592 \\
        \bottomrule
    \end{tabular}
\end{table}

\newpage
\setcounter{table}{3}
\begin{table}[h]  
    \centering
    \caption{Distribution of impactor core material in all iSALE simulations (continued, small-size impactors).}
    \label{tab:Results_Large_3}
    \begin{tabular}{c|c|c|cccc}
        \toprule
        Impactor Mass & Impact Velocity & Impact Angle  &  Escape  &  Core   &  Mantle  &  Vapor \\
        (kg)          & (km s$^{-1}$)   & (degrees)     &  \quad   &  \quad  &  \quad   &  \quad \\
        \midrule
            \multirow[c]{26}{*}{2.78e21} & \multirow[c]{8}{*}{20.0} & 00 & 0.00000 & 0.97222 & 0.02778 & 0.00000 \\
            \quad & \quad & 30 & 0.00000 & 0.00000 & 1.00000 & 0.00000 \\
            \quad & \quad & 45 & 0.00000 & 0.00000 & 0.95833 & 0.04167 \\
            \quad & \quad & 50 & 0.00000 & 0.00000 & 0.86574 & 0.13426 \\
            \quad & \quad & 55 & 0.02778 & 0.00000 & 0.70833 & 0.26389 \\
            \quad & \quad & 60 & 0.41204 & 0.00000 & 0.13889 & 0.44907 \\
            \quad & \quad & 65 & 0.50463 & 0.00000 & 0.01852 & 0.47685 \\
            \quad & \quad & 70 & 0.89815 & 0.00000 & 0.00463 & 0.09722 \\
            \quad & \multirow[c]{10}{*}{16.0} & 00 & 0.00000 & 0.99074 & 0.00926 & 0.00000 \\
            \quad & \quad & 30 & 0.00000 & 0.00000 & 0.99074 & 0.00926 \\
            \quad & \quad & 35 & 0.00000 & 0.00000 & 1.00000 & 0.00000 \\
            \quad & \quad & 40 & 0.00000 & 0.00000 & 0.99537 & 0.00463 \\
            \quad & \quad & 45 & 0.00000 & 0.00000 & 0.99074 & 0.00926 \\
            \quad & \quad & 50 & 0.00000 & 0.00000 & 0.95370 & 0.04630 \\
            \quad & \quad & 55 & 0.00000 & 0.00000 & 0.83796 & 0.16204 \\
            \quad & \quad & 60 & 0.10185 & 0.00000 & 0.38426 & 0.51389 \\
            \quad & \quad & 65 & 0.34259 & 0.00000 & 0.28704 & 0.37037 \\
            \quad & \quad & 70 & 0.79630 & 0.00000 & 0.00463 & 0.19907 \\
            \quad & \multirow[c]{8}{*}{11.2} & 00 & 0.00000 & 0.99074 & 0.00926 & 0.00000 \\
            \quad & \quad & 30 & 0.00000 & 0.00000 & 1.00000 & 0.00000 \\
            \quad & \quad & 45 & 0.00000 & 0.00000 & 1.00000 & 0.00000 \\
            \quad & \quad & 50 & 0.00000 & 0.00000 & 0.99074 & 0.00926 \\
            \quad & \quad & 55 & 0.00000 & 0.00000 & 0.89352 & 0.10648 \\
            \quad & \quad & 60 & 0.00000 & 0.00000 & 0.81019 & 0.18981 \\
            \quad & \quad & 65 & 0.00000 & 0.00000 & 0.63889 & 0.36111 \\
            \quad & \quad & 70 & 0.00000 & 0.00000 & 0.67593 & 0.32407 \\
        \bottomrule
    \end{tabular}
\end{table}

\end{document}